\documentclass [glov3] {svjour3}

\usepackage{float}
\usepackage{pgf-pie}
\usepackage{amsmath,amsfonts}
\usepackage{algorithmic}
\usepackage{graphicx}
\usepackage{textcomp}
\usepackage{xcolor}
\usepackage{enumitem}
\usepackage{listings}
\usepackage{xcolor}
\usepackage{microtype}
\usepackage{natbib}
\usepackage{wheelchart}
\usepackage{academicons}
\usepackage[T1]{fontenc}
\usepackage[hidelinks]{hyperref}
\definecolor{orcidlogocol}{HTML}{A6CE39}

\setcitestyle{authoryear,open={(},close={)}} 
\definecolor{codegreen}{rgb}{0,0.6,0}
\definecolor{codegray}{rgb}{0.5,0.5,0.5}
\definecolor{codepurple}{rgb}{0.58,0,0.82}
\definecolor{backcolour}{rgb}{0.95,0.95,0.92}

\lstdefinestyle{mystyle}{
    backgroundcolor=\color{backcolour},   
    commentstyle=\color{codegreen},
    keywordstyle=\color{magenta},
    numberstyle=\tiny\color{codegray},
    stringstyle=\color{codepurple},
    basicstyle=\ttfamily\footnotesize,
    breakatwhitespace=false,         
    breaklines=true,                 
    captionpos=b,                    
    keepspaces=true,                 
    numbers=left,                    
    numbersep=5pt,                  
    showspaces=false,                
    showstringspaces=false,
    showtabs=false,                  
    tabsize=2
}
\AtBeginDocument{%
  \providecommand\BibTeX{{%
    \normalfont B\kern-0.5em{\scshape i\kern-0.25em b}\kern-0.8em\TeX}}}

\begin{document}

    \title {Evaluating the Code Quality of AI-Assisted Code Generation Tools: An Empirical Study on GitHub Copilot, Amazon CodeWhisperer, and ChatGPT}
    
    \author {Burak Yetiştiren \and Işık Özsoy \and Miray Ayerdem \and Eray Tüzün}
    \institute{Burak Yetiştiren \at Bilkent University,\\\email{burakyetistiren@hotmail.com}
    \and Işık Özsoy \at Bilkent University,\\\email{ozsoyisik@gmail.com
    \and Miray Ayerdem \at Bilkent University,\\\email{miray.ayerdem@ug.bilkent.edu.tr
    \and Eray Tüzün \at Bilkent University,\\\email{eraytuzun@cs.bilkent.edu.tr}}}}
    \maketitle

    \begin{abstract}
        
        \hfill \break
        \begin{itemize}[leftmargin=0cm]
            \item[] \textbf{Context} AI-assisted code generation tools have become increasingly prevalent in software engineering, offering the ability to generate code from natural language prompts or partial code inputs. Notable examples of these tools include GitHub Copilot, Amazon CodeWhisperer, and OpenAI's ChatGPT. \\

            \item[] \textbf{Objective} This study aims to compare the performance of these prominent code generation tools in terms of code quality metrics, such as Code Validity, Code Correctness, Code Security, Code Reliability, and Code Maintainability, to identify their strengths and shortcomings.\\

            \item[] \textbf{Method} We assess the code generation capabilities of GitHub Copilot, Amazon CodeWhisperer, and ChatGPT using the benchmark HumanEval Dataset. The generated code is then evaluated based on the proposed code quality metrics.
            \\
            \item[] \textbf{Results} Our analysis reveals that the latest versions of ChatGPT, GitHub Copilot, and Amazon CodeWhisperer generate correct code 65.2\%, 46.3\%, and 31.1\% of the time, respectively. In comparison, the newer versions of GitHub CoPilot and Amazon CodeWhisperer showed improvement rates of 18\% for GitHub Copilot and 7\% for Amazon CodeWhisperer. The average technical debt, considering code smells, was found to be 8.9 minutes for ChatGPT, 9.1 minutes for GitHub Copilot, and 5.6 minutes for Amazon CodeWhisperer. \\
            
            \item[] \textbf{Conclusions} This study highlights the strengths and weaknesses of some of the most popular code generation tools, providing valuable insights for practitioners. By comparing these generators, our results may assist practitioners in selecting the optimal tool for specific tasks, enhancing their decision-making process. 
            
        \end{itemize}
    \end{abstract}
    
    

    \section{Introduction}\label{Introduction}

       Code completion and generation tools are essential for enhancing programmers' performance and output quality in software development.  \cite{omar2012active} define code completion tools as tools that are offered in most editors, which list contextually-relevant variables, fields, methods, types, and other code snippets in the form of a floating menu. By exploring and making choices from this menu, developers can avoid frequent grammatical and logical errors, reduce redundant keystrokes, and explore new APIs without having to go through the mental effort of switching to an external documentation tool or API browser. Some of the well-known code completion tools include IntelliSense in Visual Studio Code\footnote{\href{https://code.visualstudio.com/docs/editor/intellisense}{code.visualstudio.com/docs/editor/intellisense}} and the built-in code completion in the JetBrains IDEs\footnote{\href{https://www.jetbrains.com/}{jetbrains.com}}. Although these tools can output code snippets, they differ fundamentally from code generators.
        
       The advent of advanced language processing technologies has led to the emergence of Large Language Models (LLMs). While LLMs have numerous use cases, we focus on their code generation capabilities. Unlike code completion tools, code generators actively utilize LLMs by providing the programmer's input to the specified LLM and returning the output to the programmer's workspace. Currently, code generators' outputs cannot be produced locally, unlike code completion tools. Additionally, code generators can generate longer outputs in the form of lines or blocks of code, which can build function bodies or other constructs. Moreover, code generators can convert natural language inputs into source code, a key distinction from code completion tools.
        
       Our motivation for conducting this study stems from the growing interest in AI-assisted code generators and the spread of unverified information about them. We recognize the popularity and potential of state-of-the-art AI-assisted code generators, as well as the heuristic feedback from various communities. In line with our previous study \citep{yetistiren2022assessing}, we believe it is worthwhile to evaluate the potential benefits of these generators. Although these tools can generate code, their value remains undetermined. To systematically assess these code generators, we propose an experimental setup evaluating the generated code based on Code Validity, Code Correctness, Code Security, Code Reliability, and Code Maintainability.

        We have chosen GitHub Copilot, Amazon CodeWhisperer, and ChatGPT for our study, leading us to formulate the following research questions:
               
        \begin{itemize}[leftmargin=0cm]
            \item[] \textbf{RQ1} What is the quality of the code generated by the code generation tools? 
                \subitem \textbf{RQ1.1} How valid are the code generation tools’ code suggestions? 
                \subitem \textbf{RQ1.2} How correct are code generation tools’ code suggestions?
                \subitem \textbf{RQ1.3} How secure are code generation tools’ code suggestions?
                \subitem \textbf{RQ1.4} How reliable are code generation tools’ code suggestions?
                \subitem \textbf{RQ1.5} How maintainable are code generation tools’ code suggestions?
            \item[] \textbf{RQ2} What is the impact of using the docstrings on the generated code quality?
            \item[] \textbf{RQ3} What is the impact of using meaningful function names on the generated code quality?
            \item[] \textbf{RQ4} How did the code generation tools evolve over time?
        \end{itemize}
        
      We believe our study will enable users to more effectively leverage AI-assisted code generators for generating accurate, valid, reliable, maintainable, and secure results. In addition, tool developers can benefit from our findings to identify and enhance the strengths and address the weaknesses of their tools in real-world situations. The comparative aspect of our study provides valuable insights into the performance of each code generation tool relative to its competitors.

        The structure of our study is as follows: In Section \ref{Background}, we provide some background information about the code generation tools we evaluate. In Section \ref{Methodology}, we provide a detailed explanation of the research questions we have determined by elaborating on our experimental setup. Our results are presented in Section \ref{Results}, and they are discussed in Section \ref{Discussion}. The threats that influence the validity of our study are addressed in Section \ref{Threats-to-Validity}. In Section \ref{Related-Work}, we discuss related work. Finally, in Section \ref{Conclusion}, we conclude our study.

    \section{Background}\label{Background}
        \begin{table}
        {\renewcommand{\arraystretch}{1.25}%
            \centering
            \scalebox{0.80}{%
            \begin{tabular}{ 
        |p{3.5cm}||p{3.5cm}|p{3.5cm}|p{3.5cm}|}
                Features &  ChatGPT & Amazon CodeWhisperer & GitHub Copilot\\
                 \hline

               IDE Support  & No IDE Support  & JetBrains, Visual Studio Code, AWS Cloud9, or the AWS Lambda console & IntelliJ IDEA, Android Studio, AppCode, CLion, Code With Me Guest, DataGrip, DataSpell, GoLand, JetBrains Client, MPS, PhpStorm, PyCharm, Rider, RubyMine, WebStorm
 \\
             \hline
               First Release Time & Nov-30-2022 & June-23-2022 & Oct-29-2021 \\
                \hline
               Developer & OpenAI & AWS & OpenAI-Microsoft \\
               \hline
               Providing References to Suggestions & NO & YES & NO \\
                 \hline
               Explanation of Suggestions & YES & NO & NO \\
                \hline
               Providing Multiple Suggestions & NO (Theoretically user can manually ask for another suggestion.) & YES (Up to 5) & YES (Up to 10)\\
              
                \hline
               Training Data Source &  

GitHub Repositories, \newline OpenAI Codex Dataset, other code repositories such as GitLab, Bitbucket, and SourceForge  &``Vast amounts of publicly available code" &``...trained on all languages that appear in public repositories" (Fine-tuned) \\ 
 \hline
Programming Languages work best with (according to the vendor) & N/A & C\#, Java, JavaScript, Python, and TypeScript & C, C++, C\#, Go, Java, JavaScript, PHP, Python, Ruby, Scala, and TypeScript\\
 \hline
               Multipurpose (other than programming) & YES & NO & NO \\
                \hline
               Subscription & ChatGPT Free  \newline ChatGPT Plus (\$20 per month) & Free Preview & Copilot for Students (Free)\newline Copilot for Individuals (\$10 per month)\newline Copilot for Business (\$19 per user, per month) \\
                \hline
               Can be Used Offline? & NO & NO & NO \\

                \hline
               Can it Access Local Files? & NO & YES & YES \\
                \hline

            \end{tabular}}
            \caption{Comparing Relevant Code Generation Tools}
            \label{tab:comparison-table}
        }\end{table}
        \subsection{GitHub Copilot}\label{Background-GitHub-Copilot}
            GitHub Copilot\footnote{\href{https://github.com/features/copilot/}{copilot.github.com}} is a code generation tool that utilizes a variety of technologies, including a compatible IDE, and the OpenAI Codex Model\footnote{\href{https://openai.com/blog/openai-codex/}{openai.com/blog/openai-codex}}. GitHub announced GitHub Copilot for technical preview in the Visual Studio Code development environment on June 29, 2021 \citep{friedman2021introducing}. GitHub declared on June 21, 2022, that Copilot was out of the technical preview phase and is now accessible as a subscription-based service for individual developers \citep{dohmke2022github}. It currently has subscription plans for individuals and businesses. GitHub Copilot can be installed and used as an extension to Visual Studio Code, Neovim, IDEs developed by JetBrains\footnote{\href{https://plugins.jetbrains.com/plugin/17718-github-copilot}{plugins.jetbrains.com/plugin/17718-github-copilot}}, and GitHub Codespaces\footnote{\href{https://github.com/features/codespaces}{github.com/features/codespaces}}. The underlying service continuously takes user code samples and sends the snippets to the underlying OpenAI Codex Model. GitHub Copilot generates the code and presents the results of the OpenAI Codex Model by adjusting the generated code to the current workspace of the programmer \citep{ernst2022ai}. 

            The Codex model relies on Generative Pre-trained Transformer (GPT) models the company previously invented for text generation. The public code available on GitHub was used during the fine-tuning of the model to implement the code recognition and generation capabilities.
            
        \subsection{Amazon CodeWhisperer}\label{Background-Amazon-CodeWhisperer}
            Amazon CodeWhisperer\footnote{\href{https://aws.amazon.com/codewhisperer/}{aws.amazon.com/codewhisperer}} improves developer productivity by generating code recommendations based on both developers' comments in English and prior code in the IDE. AWS announced Amazon CodeWhisperer Preview on June 23, 2022, \citep{bays2022AWS}. The code recommendations provided by CodeWhisperer are based on ML models trained on various data sources, such as Amazon's sources and other open-source codes. When developers write a comment in their IDE's code editor, CodeWhisperer will automatically examine the comment and determine the best-suited cloud services and public libraries. Then, it will provide a code snippet directly within the code editor. Moreover, CodeWhisperer simplifies the use of AWS services for developers by offering suggestions for AWS API code across top services such as Amazon Elastic Compute Cloud (EC2), AWS Lambda, and Amazon Simple Storage Service (S3).
            
            CodeWhisperer supports multiple IDEs including JetBrains, Visual Studio Code, AWS Cloud9, or the AWS Lambda console as part of the AWS IDE toolkit. Moreover, it currently supports Java, JavaScript, Python, C\#, and Typescript. As an additional feature, CodeWhisperer has a reference tracker that detects the code recommendations similar to particular CodeWhisperer training data and provides those references to developers. CodeWhisperer can also scan the code and define the security issues.  
        \subsection{ChatGPT}\label{Background-ChatGPT}
            ChatGPT\footnote{\href{https://openai.com/blog/chatgpt}{openai.com/blog/chatgpt}} is a language model announced by OpenAI on November 30, 2022. The subscription plan, ChatGPT Plus, is available since February 1, 2023 \citep{openai2023Introducing}.  ChatGPT uses advanced machine learning algorithms to generate human-like text responses. It is trained on vast amounts of text data from the internet. It is capable of answering a wide range of questions, admitting its mistakes, challenging incorrect premises, and rejecting inappropriate requests. While the primary purpose of a chatbot is to imitate human conversation, ChatGPT is highly versatile and can perform a wide range of tasks such as coding and debugging software, providing responses to exam questions, composing poetic works and musical lyrics, and more \citep{tung2023Chatgpt}.
            
            ChatGPT has been adjusted specifically from a model within the GPT-3.5 series\footnote{\href{https://platform.openai.com/docs/model-index-for-researchers}{platform.openai.com/docs/model-index-for-researchers}}, which completed its training process early in 2022 using supervised learning as well as reinforcement learning. Moreover, OpenAI continues to gather information from ChatGPT users to improve and refine its performance.  

            It is also notable that ChatGPT has become the fastest-growing app in history according to the study of the Union Bank of Switzerland (UBS). In January 2023, ChatGPT attracted 13 million unique visitors daily, over twice the number it received in December according to the study. The report also states that despite being only two months old, ChatGPT has already reached a monthly user base of 100 million. \citep{cerullo2023Chatgpt}.

    \subsection{Comparison of ChatGPT, GitHub Copilot, and Amazon CodeWhisperer}\label{Background-Comparison-Tools}
        When we performed the experiment on GitHub Copilot, Amazon CodeWhisperer, and ChatGPT for our study, we could observe some advantages and limitations of the tools. According to our observations, and the background knowledge we stated above, we created Table \ref{tab:comparison-table}. In the table, it can be seen that while GitHub Copilot and Amazon CodeWhisperer have IDE support, ChatGPT does not have this yet, apart from its API support. Moreover, as we mentioned in earlier parts of Section \ref{Background}, ChatGPT and Amazon CodeWhisperer were introduced in 2022. However, GitHub Copilot was announced in 2021. Hence, it can be said that GitHub Copilot is one of the pioneers of this field.  Additionally, it is notable that OpenAI developed both ChatGPT and GitHub Copilot while AWS developed Amazon CodeWhisperer. ChatGPT does not have any specific information about supported programming languages but GitHub Copilot and Amazon CodeWhisperer specify the supported programming languages and we can see that GitHub Copilot supports more programming languages than Amazon CodeWhisperer in Table \ref{tab:comparison-table}. Although Amazon CodeWhisperer is still free as it is in the technical preview stage, and GitHub Copilot has different subscription plans. ChatGPT is also available in the technical preview but it also has a subscription plan.

        Furthermore, we added our observations about the tools to Table \ref{tab:comparison-table}. Firstly, we observed that Amazon CodeWhisperer and GitHub Copilot could provide more than one recommendation and it would be easy to choose the most relevant one from the options for users. By contrast, ChatGPT mainly provided one suggestion at a time unless we did not ask for additional suggestions. On the other hand, ChatGPT was the only tool that explained its recommendations in detail and was used for purposes other than programming. We also observed Amazon CodeWhisperer was the only tool that presented the recommendations' source. It is notable to mention that while Amazon CodeWhisperer and GitHub Copilot could access users' local files, ChatGPT could not access them. Lastly, we observed that none of these tools could be used offline.

    \section{Methodology}\label{Methodology}

        Under the subsections below, we elaborate on our methodology. Section \ref{Methodology-HumanEval-Dataset} gives detail about the data we use. To address the research questions, we created an experimental setup, which systematically evaluates the effectiveness of code generation tools, that is described in Section \ref{Methodology-Experimental-Setup}. The details of our assessment are presented in Section \ref{Methodology-Our-Metrics}. In Sections \ref{Methodology-Using-only-Function-Signatures} and \ref{Methodology-Using-Dummy-Function-Names}, we elaborate on the two additional experiments we conducted to test the effect of the function names and explanations of the generated code quality. Moreover, we use different versions of GitHub Copilot and Amazon CodeWhisperer to assess the performance of those code generation tools over time, which is described in Section \ref{Methodology-Evaluation-of-Code-Generation-Tools-Over-Time}.

        \begin{figure}[h]
            \centering
            \includegraphics[scale=0.5]{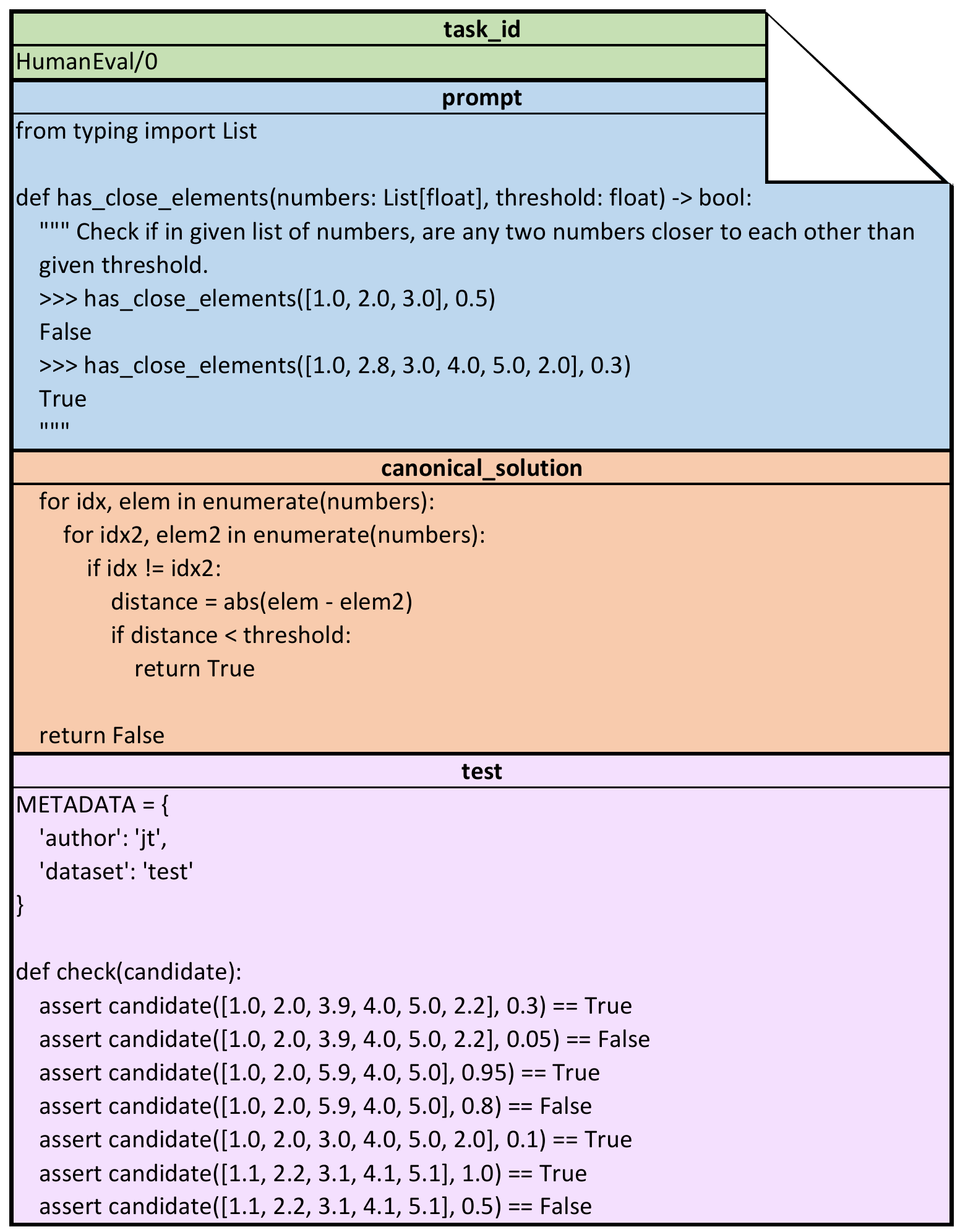}
            \caption{Example Problem (ID: 0) from HumanEval dataset}
            \label{example_problem}
        \end{figure}
    
        \begin{figure}[h]
\begin{lstlisting}[caption={Generated Code for the Example Problem by GitHub Copilot \textit{v1.7.4421} (ID: 0)}, label={prompt-0}, numbers=none]
from typing import List

def has_close_elements(numbers: List[float], threshold: float) -> bool:
    """ Check if in given list of numbers, are any two numbers closer to each other than given threshold.
    >>> has_close_elements([1.0, 2.0, 3.0], 0.5)
    False
    >>> has_close_elements([1.0, 2.8, 3.0, 4.0, 5.0, 2.0], 0.3)
    True
    """
    for i in range(len(numbers)):
        for j in range(i + 1, len(numbers)):
            if abs(numbers[i] - numbers[j]) < threshold:
                return True
    return False
\end{lstlisting}
        \end{figure}
        
        \subsection{HumanEval Dataset}\label{Methodology-HumanEval-Dataset}
            For our experiment, we use the HumanEval dataset proposed by \cite{copilot2021}. This dataset contains 164 problems. Each problem is accompanied by a task ID, a prompt, the canonical solution, and unit tests. The structure of a problem can be viewed in Figure \ref{example_problem}. The task ID is the ID of that particular problem which ranges from 0 to 163. The prompt part contains the function prototype, the explanation of the problem, some function calls and their output in a Python docstring, and library imports, if applicable. A canonical solution is considered as a correct solution which is coded by a ``human'' programmer. The test part contains unit tests as a Python function. 
    
            We pass the function prototype and the docstring as input to code generation tools. An example code generation done by GitHub Copilot, where the input problem is shown in Figure \ref{example_problem} can be viewed in Listing \ref{prompt-0}.

        \subsection{Experimental Setup}\label{Methodology-Experimental-Setup}

            \begin{figure*}
                \includegraphics[scale=0.43]{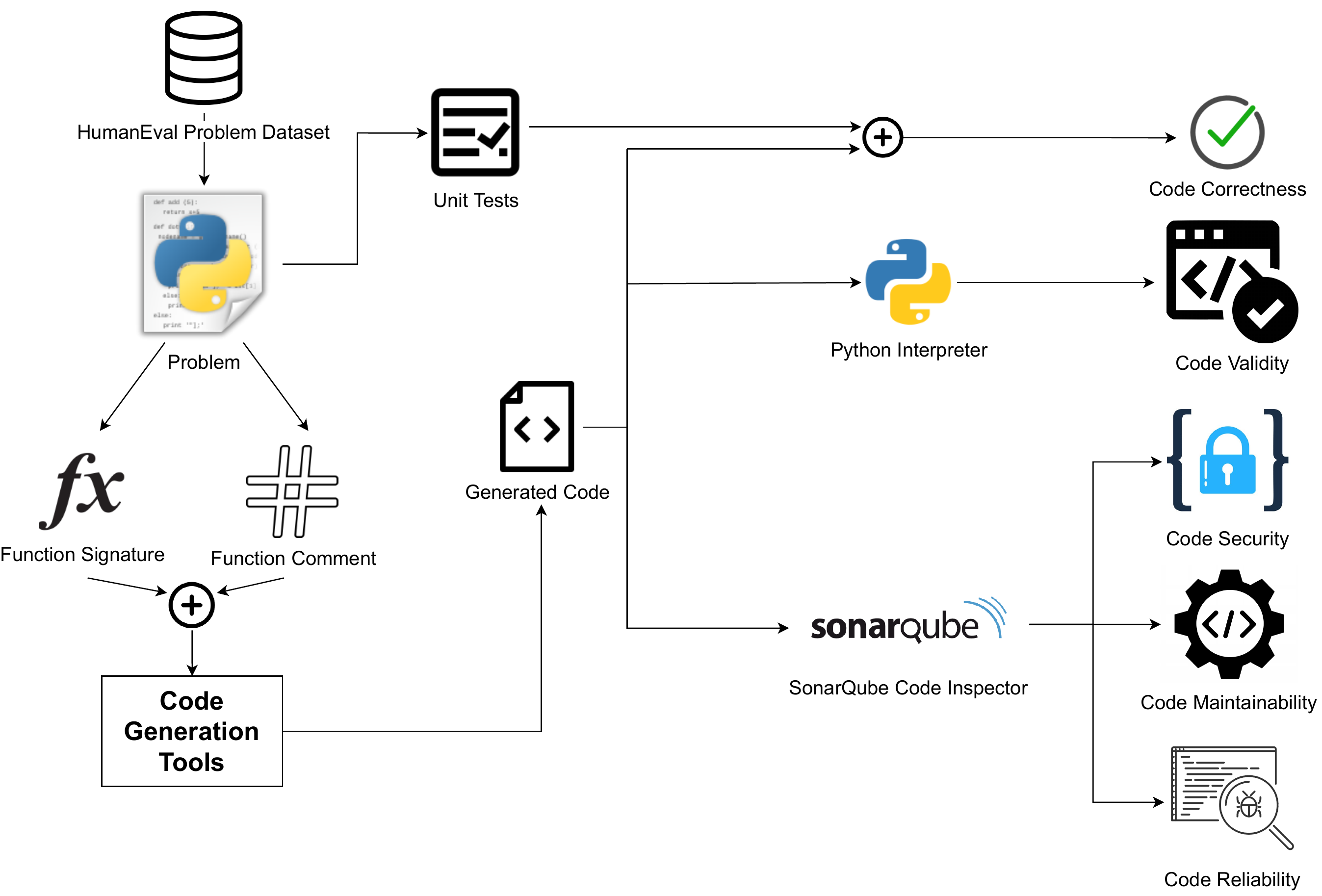}
                \caption{Experimental Setup}
                \label{experimental_setup_fig}
            \end{figure*}

            \begin{figure*}
              \includegraphics[scale=0.57]{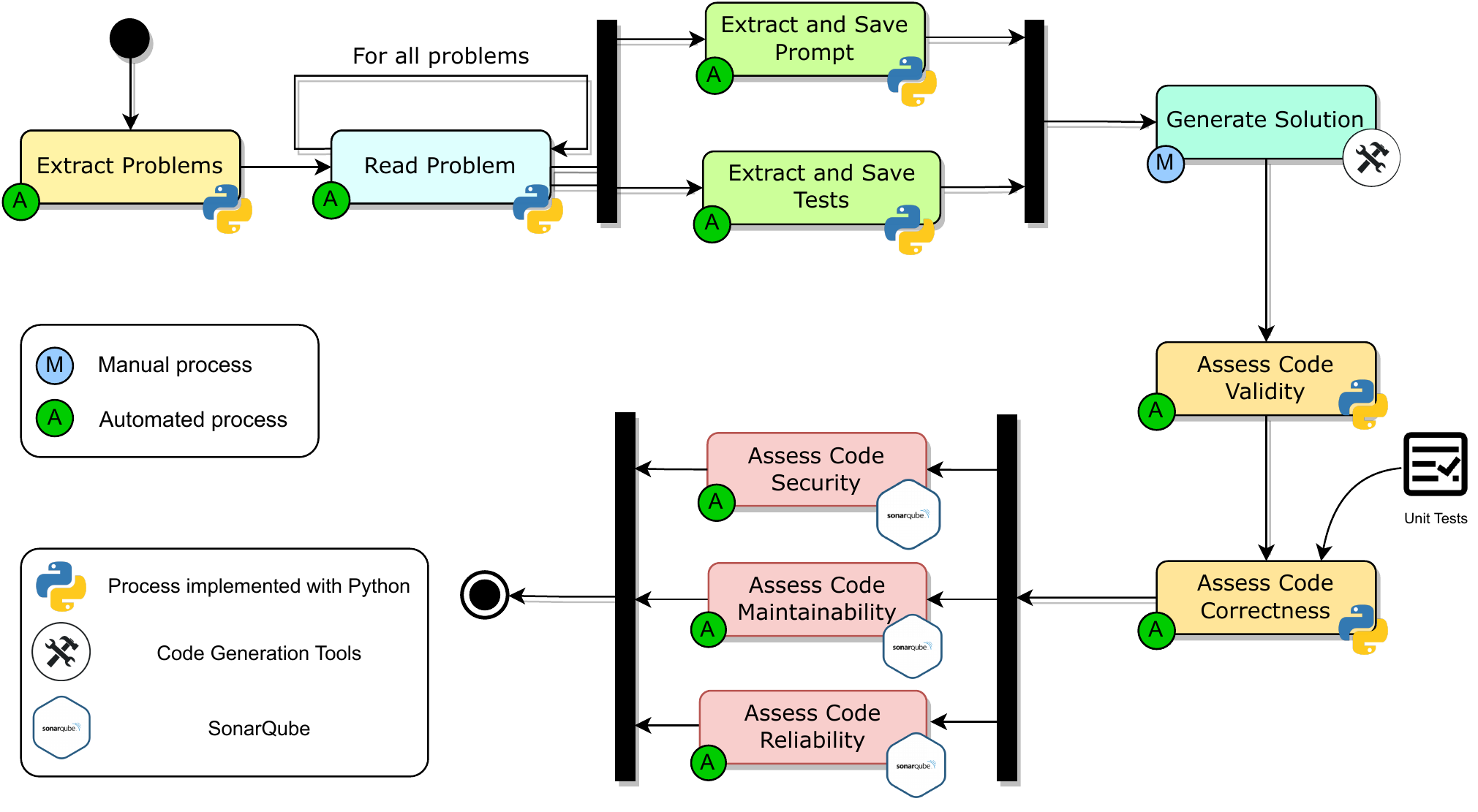}
              \caption{Experiment Workflow}
              \label{experimental_workflow}
            \end{figure*}

            In Figure \ref{experimental_setup_fig}, we focus on what artifacts were employed for which tools, and which metric these combinations correspond to. Whereas in Figure \ref{experimental_workflow}, we provide a step-by-step illustration of the experiment's workflow. In this figure, given the HumanEval problem dataset \citep{copilot2021}, we start our experiment by extracting the problems. We achieve this by reading the dataset and representing each problem with a separate JSON format file. After completing the extraction procedure, we save the unit tests and the prompt of a problem as separate Python files to the directory corresponding to the problem's ID. Subsequently, we generate solutions using an already prepared Python file containing the prompt. This prompt combines the function signature and docstring contained in the function body. Given the dynamic characteristic of code generation tools, GitHub Copilot, Amazon CodeWhisperer, and ChatGPT, in terms of the interactions between the programmer and the service, we implement the code generation step of our experiment manually by employing the Visual Studio Code IDE for GitHub Copilot and Amazon CodeWhisperer and using the interface provided by OpenAI in a given browser for ChatGPT. After the code generation step is completed, we start the assessment phase by executing the tests on the generated solutions to assess code validity and code correctness. After this, we utilize SonarQube to find the security rating, the number of bugs, and the number of code smells for each problem, these correspond to the Code Security, Code Reliability, and Code Maintainability metrics. We detect the bugs and code smells separately, since the code smells are mostly different than bugs, they do not necessarily cause the code to be incorrect, but introduce some uneasiness in the code which can cause more problems in the future. For each step of the assessment phase, we save the individual assessment results related to the problem. The extracted results can be seen in our reproduction package\footnote{\href{https://github.com/mirayayerdem/Github-Copilot-Amazon-Whisperer-ChatGPT/blob/main/misc/All_Experiment_Results.xlsx}{https://github.com/mirayayerdem/Github-Copilot-Amazon-Whisperer-ChatGPT/blob/main/misc/All\_Experiment\_Results.xlsx}}. 
    
            

            
            We further test the validity of our findings in our literature survey about providing code generation tools with code comments and function signatures, by implementing two additional experiments about the significance of function names, parameters, and comments explained in Sections \ref{Methodology-Using-only-Function-Signatures} and \ref{Methodology-Using-Dummy-Function-Names}. 
            

        \subsection{Code Metrics}\label{Methodology-Our-Metrics}
            We have evaluated our results in terms of code validity, code correctness, code security, code reliability, and code maintainability. Our metric for code validity is binary, in which we have two possible values, `0' and `1' indicating if the solution is valid or not. This is assessed in terms of how a given code segment is compliant with the rules and regulations (i.e., syntax rules) of a given programming language and with any errors that could be raised during runtime. The dataset we use is constructed for the Python programming language; therefore, to check for code validity, we use the Python 3.10.10 interpreter. 
            
    
            For code correctness, we want to assess the extent to which the generated code performs as intended. As we previously stated, the problems in the HumanEval dataset are accompanied by problem-specific unit tests. On average, each problem comes with 7.7 unit tests \citep{copilot2021}. We measured the code correctness as passed unit tests divided by all unit tests for a specific problem. Considering the abundance of unit tests, we believe that the most convenient way to assess code correctness is to make use of the provided tests.

            We have also evaluated the average code correctness which is measured as the sum of all code correctness scores divided by the problem count. While calculating average code correctness, we consider the code correctness score of invalid code generations as 0.

            We show our calculation methods for Code Correctness and Average Code Correctness below. $CCS$ stands for Code Correctness Score, and $CCS_{i}$ is the Code Correctness Score for the $i\textsuperscript{th}$ problem. The range of $i$ is 0 to 163.
            $$
            Code\:Correctness = \frac{\sum_{i=0}^{163}{CCS_{i}\:[CCS_{i}=1]}}{164}
            $$
            \\
            $$
            Average\:Code\:Correctness = \frac{\sum_{i=0}^{163}{CCS_{i}}}{164} 
            $$
            \\
            
            Finally, we used SonarQube\footnote{\href{https://www.sonarqube.org}{sonarqube.org}} to assess code security, code reliability, and code maintainability metrics. For code security, we define the term vulnerability, which compromises the security of the code; therefore, introducing security risks, considering the possible deployment of the code in question as software or part of the software. We use SonarQube's Security Module to assess code security. This module calculates the number of vulnerabilities in a given code. To assess code maintainability, SonarQube runs its evaluation on the given code in terms of the count of code smells present in the code. For code reliability, we count the number of bugs in the code using SonarQube.
        
        \subsection{Using only Function Signatures (RQ2)}\label{Methodology-Using-only-Function-Signatures}
            In this experiment, we removed the docstrings from the problems to assess the effect of docstrings on the generated solution. The docstring of a given problem in the HumanEval dataset includes the explanation of the function as the intended purpose of what that problem should be doing. This explanation is then accompanied by some sample test cases and their results (an example can be seen in the ``prompt'' part in Figure \ref{example_problem}). We used GitHub Copilot, Amazon CodeWhisperer, and ChatGPT to generate code by only using the name and the parameters of the function as a reference. We aimed to see how our results would change in comparison to our previous results.
            
        \subsection{Using Dummy Function Names (RQ3)}\label{Methodology-Using-Dummy-Function-Names}
            We changed the function names of the problems with the dummy function name `foo', to assess the effect of meaningful function names on the generated solution. The tools are then employed to generate code with such inputs. We assess the generated code using the Code Validity and Correctness metrics.

        \subsection{Evaluation of Code Generation Tools Over Time (RQ4)}\label{Methodology-Evaluation-of-Code-Generation-Tools-Over-Time}
        Since the initial release of GitHub Copilot, there were multiple official updates that the tool received, apart from the continuous training of the underlying LLM of GitHub Copilot. Considering that we have ready-to-use results for GitHub Copilot from our old study \citep{yetistiren2022assessing}, as a part of this study we are also evaluating how a given code generation tool has evolved over time. In that regard, we will be comparing GitHub Copilot \textit{v1.7.4421} and \textit{v1.70.8099} versions. Since AWS does not specify the version of CodeWhisperer, we are unable to specify the versions we have used to run the prior and later experiments; we can only provide the dates of our two experiments, which are November `22 for prior and January `23 for the latter experiment. For this research question, we use the Code Correctness and Code Validity metrics for the three experiment types we explained in Sections \ref{Methodology-Experimental-Setup}, \ref{Methodology-Using-only-Function-Signatures}, and \ref{Methodology-Using-Dummy-Function-Names}.

    \section{Results}\label{Results}
        \subsection{Code Validity (RQ1.1)}\label{Results-Validity}

        The use of interpreters in Python made it easier for us to evaluate code validity by simply trying to execute the code and catch errors at runtime. This should not be confused with runtime errors; in Python, like runtime errors, the syntax errors can also be detected by executing the script, unlike the compiler approach used in other high-level programming languages like Java and C++. 

        As we noted earlier, our metric for code validity is binary, such that if any errors were raised during the execution of a given Python script, we denoted that script as invalid. Moreover, for such scripts, we did not calculate the correctness score, as consideration of such scores could impose possible threats to the validity of our evaluation. 

\begin{figure*}
                \includegraphics[scale=0.6]{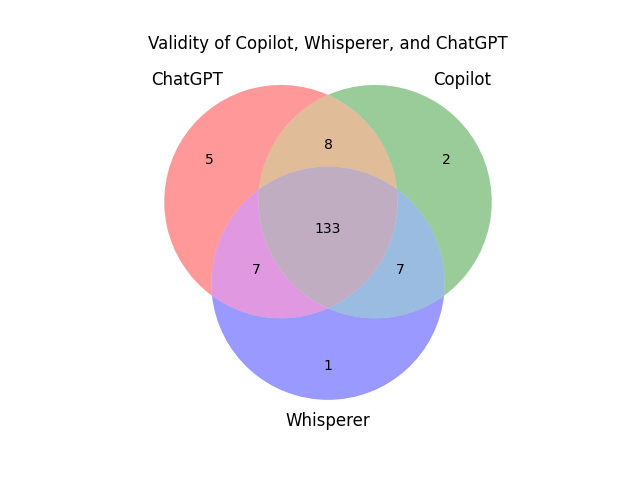}
                \caption{Distribution of Validly Generated Samples among the Code Generation Tools}
                \label{fig:code_validity_scores}
            \end{figure*}

        The full Code Validity results of our experiments can be visualized in Figure \ref{fig:code_validity_scores}. Out of 164 generations of GitHub Copilot to the problems, 150 were valid. This yielded a \begin{math} 91.5\% \end{math} success rate in terms of generating valid code. Amazon CodeWhisperer generated a valid code for 148 of the problems, which yields a \begin{math} 90.2\% \end{math} success rate. ChatGPT was able to generate a valid code for 153 problems, which yields a \begin{math} 93.3\% \end{math} success rate.

        \subsection{Code Correctness (RQ1.2)}\label{Results-Code-Correctness}

        \begin{figure*}
            \includegraphics[scale=0.50]{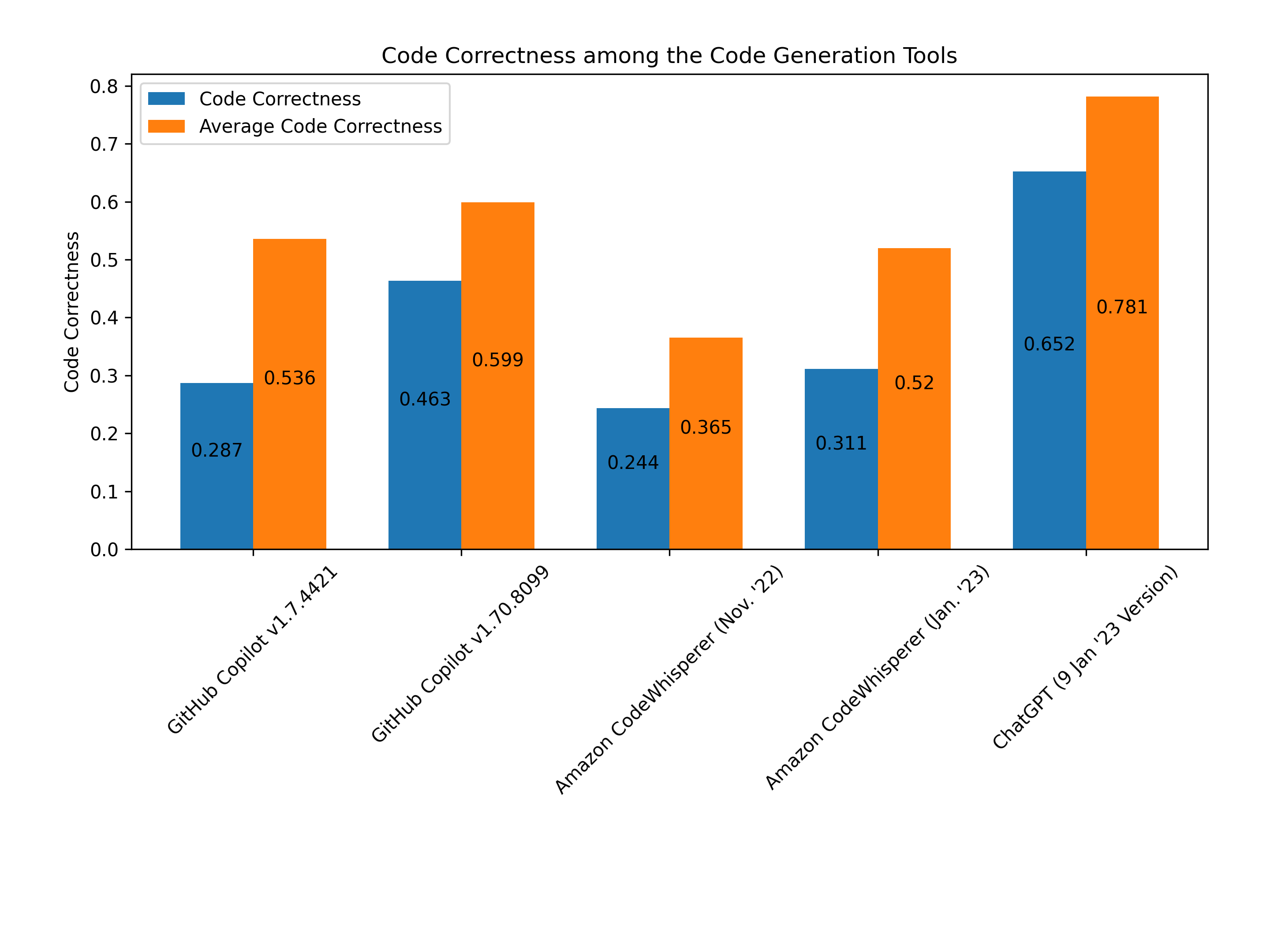}
            \caption{Code Correctness and Average Code Correctness Scores of the Code Generation Tools}
            \label{result_average_correctness}
        \end{figure*}

        We used the number of passed unit tests divided by all unit tests to calculate the success percentage of the code for each problem, which we have defined in Section \ref{Methodology-Our-Metrics}. In Figures \ref{wheel_chart_copilot_a} - \ref{wheel_chart_chatgpt_b}, we provided the percentage distribution of code generations falling under different categories (correct, partially correct, and incorrect). Moreover, we also measured the average code correctness score by dividing the summation of all code correctness scores by the number of all problems, which we have again defined in Section \ref{Methodology-Our-Metrics}. In Figure \ref{result_average_correctness}, we provided the comparisons of code correctness scores and average code correctness scores among code generation tools.

        \begin{minipage}{\linewidth}
            \wheelchartcopilota{
                46.3/A0/{Proportion of Correct\newline Generations},
                23.2/C0/{Proportion of Partially\newline Correct Generations},
                30.5/B0/{Proportion of Incorrect\newline Generations}
            }
        \end{minipage}
            
        \begin{minipage}{\linewidth}
            \wheelchartcopilotb{
                26.3/A0/\begin{math} 100\% > CCS_{i} > 75\% \end{math},
                34.2/B0/\begin{math} 75\% \geq{CCS_{i}} > 50\%  \end{math},
                23.7/C0/\begin{math} 50\% \geq{CCS_{i}} > 25\%   \end{math},
                15.8/D0/\begin{math} 25\% \geq{CCS_{i}} > 0\%  \end{math}
            }
        \end{minipage}

        We observed that for \begin{math} 46.3\% \end{math} of the problems, GitHub Copilot managed to generate the correct code for the given problem, whereas it completely failed to provide a correct solution for \begin{math} 30.5\% \end{math} of the problems. Generated solutions for the remaining \begin{math} 23.2\% \end{math} of the problems were partially correct, as shown in Figure \ref{wheel_chart_copilot_a}. Partially correct generations are the ones that pass at least one of the unit tests but not all of them. We believe partially correct generations are useful, with the assumption that if at least one unit test is passing, this is a potential indicator that with further improvements by the programmer, the code could become correct. To analyze the partially correct code generations, we created a second pie chart in Figure \ref{wheel_chart_copilot_b}, in which we eliminated correct and incorrect code generations, yielding 38 problems. We divided \begin{math} (0, 100) \end{math} success space into four intervals. GitHub Copilot managed a success rate of \begin{math} 26.3\% \end{math} for the interval of \begin{math} 100\% > CCS_{i} > 75\% \end{math}, where \begin{math} CCS_{i} \end{math} refers to the code correctness score of the problem. Followingly, code was generated with a correctness score in the interval of \begin{math} 75\% \geq{CCS_{i}} > 50\% \end{math}, \begin{math} 34.2\% \end{math} of the time. The next interval contained the partially correct code generations with a score of \begin{math} 23.7\% \end{math}, belonging to the interval of \begin{math} 50\% \geq{CCS_{i}} > 25\%   \end{math}. For the last interval of \begin{math} 25\% \geq{CCS_{i}} > 0\%  \end{math}, the score was \begin{math} 15.8\% \end{math}. We also found the average code correctness score of GitHub Copilot as 59.85\%, shown in Figure \ref{result_average_correctness}.

        \begin{minipage}{\linewidth}
            \wheelchartwhisperera{
                31.1/A0/{Proportion of Correct\newline Generations},
                40.2/C0/{Proportion of Partially\newline Correct Generations},
                28.7/B0/{Proportion of Incorrect\newline Generations}
            }
        \end{minipage}

        \begin{minipage}{\linewidth}
            \wheelchartwhispererb{
                15.2/A0/\begin{math} 100\% > CCS_{i} > 75\% \end{math},
                37.9/B0/\begin{math} 75\% \geq{CCS_{i}} > 50\%  \end{math},
                21.2/C0/\begin{math} 50\% \geq{CCS_{i}} > 25\%   \end{math},
                25.7/D0/\begin{math} 25\% \geq{CCS_{i}} > 0\%  \end{math}
            }
        \end{minipage}

        Amazon CodeWhisperer was able to generate the correct code for \begin{math} 31.1\% \end{math} of the problems, whereas it completely failed to provide a correct solution for \begin{math} 28.7\% \end{math} of the problems. Generated solutions for the remaining \begin{math} 40.2\% \end{math} of the problems were partially correct, which is demonstrated in Figure \ref{wheel_chart_whisperer_a}. As given in Figure \ref{wheel_chart_whisperer_b} Amazon CodeWhisperer managed a success rate of \begin{math} 15.2\% \end{math} for the interval of \begin{math} 100\% > CCS_{i} > 75\% \end{math}. Followingly, code was generated with a correctness score in the interval of \begin{math} 75\% \geq{CCS_{i}} > 50\% \end{math}, \begin{math} 37.9\% \end{math} of the time. The next interval contained the partially correct code generations with a score of \begin{math} 21.2\% \end{math}, belonging to the interval of \begin{math} 50\% \geq{CCS_{i}} > 25\% \end{math}. For the last interval of \begin{math} 25\% \geq{CCS_{i}} > 0\%  \end{math}, the score was \begin{math} 25.7\% \end{math}. Additionally, we found the average code correctness score of Amazon CodeWhisperer as 51.95\%, shown in Figure \ref{result_average_correctness}.

        \begin{minipage}{\linewidth}
            \wheelchartchatgpta{
                65.2/A0/{Proportion of Correct\newline Generations},
                22.6/C0/{Proportion of Partially\newline Correct Generations},
                12.2/B0/{Proportion of Incorrect\newline Generations}
            }
        \end{minipage}

        \begin{minipage}{\linewidth}
            \wheelchartchatgptb{
                16.2/A0/\begin{math} 100\% > CCS_{i} > 75\% \end{math},
                46.0/B0/\begin{math} 75\% \geq{CCS_{i}} > 50\%  \end{math},
                24.3/C0/\begin{math} 50\% \geq{CCS_{i}} > 25\%   \end{math},
                13.5/D0/\begin{math} 25\% \geq{CCS_{i}} > 0\%  \end{math}
            }
        \end{minipage}
        \medskip

        As it can be seen in Figure \ref{wheel_chart_chatgpt_a}, ChatGPT generated the correct code for \begin{math} 65.2\% \end{math} of the problems, whereas it could not generate a correct solution for \begin{math} 12.2\% \end{math} of the problems. For the remaining \begin{math} 22.6\% \end{math} of the problems, it was able to generate partially correct code. Considering the partially correct solutions, shown in Figure \ref{wheel_chart_chatgpt_b}, ChatGPT managed a success rate of \begin{math} 16.2\% \end{math} for the interval of \begin{math} 100\% > CCS_{i} > 75\% \end{math}. Followingly, code was generated with a correctness score in the interval of \begin{math} 75\% \geq{CCS_{i}} > 50\% \end{math}, \begin{math} 46.0\% \end{math} of the time. The next interval contained the partially correct code generations with a score of \begin{math} 24.3\% \end{math}, belonging to the interval of \begin{math} 50\% \geq{CCS_{i}} > 25\%   \end{math}. For the last interval of \begin{math} 25\% \geq{CCS_{i}} > 0\%  \end{math}, the score was \begin{math} 13.5\% \end{math}. We also found the average code correctness score of ChatGPT as 78.1\%, shown in Figure \ref{result_average_correctness}.

        \subsection{Code Security \& Code Reliability \& Code Maintainability (RQ1.3 \& RQ1.4 \& RQ1.5)}\label{Results-Code-Security-Reliability-Maintainability}

        \begin{table}
          \caption{Code Security, Code Reliability, and Code Maintainability Results}
          \label{tab:results_Security-Reliability-Maintainability}
            \begin{tabular}{cccccccccc}
                {} &  \textbf{Code Security} & \multicolumn{4}{c}{\textbf{Code Reliability}} & \multicolumn{4}{c}{\textbf{Code Maintainability}}\\
                \noalign{\smallskip}\hline
                \noalign{\smallskip}
                {} & Security Rating & \multicolumn{4}{c}{Number of Bugs} & \multicolumn{4}{c}{Number of Smells}\\
                {} & $<1$ & 1 & 2 & 3 & $3<$ & 1 & 2 & 3 & $3<$ \\
                \noalign{\smallskip}\hline \noalign{\smallskip}
                Copilot v1.70.8099 (New) & 0 & 3 & 0 & 0 & 0 & 14 & 3 & 2 & 0 \\
                CodeWhisperer Jan ’23 (New) & 0 & 1 & 0 & 0 & 0 & 22 & 2 & 0 & 0 \\
                ChatGPT 9 Jan ’23 Version & 0 & 2 & 0 & 0 & 0 & 13 & 1 & 1 & 0\\
                \noalign{\smallskip}\hline
            \end{tabular}
        \end{table}

        For Code Security, Maintainability, and Reliability, we employed SonarQube to find the security rating, the number of bugs, and the number of smells for each problem. The results of our evaluation can be visualized in Table \ref{tab:results_Security-Reliability-Maintainability}. 
        
        For none of the problems for all of the generators, we did not see any security rating that was below 1, which is the maximum possible rating. Due to the constraints of the dataset we used for our study, the security results we obtained were limited. The usage of an alternative dataset with a different problem scope, and problems that yield longer solutions may reflect better Code Security results.

        Regarding Code Reliability, there were three problems containing a single bug for GitHub Copilot, one problem containing a single bug for Amazon CodeWhisperer, and two problems containing a single bug for ChatGPT. All of the bugs observed in the generations provided by Copilot (Problem IDs: \#33, \#37, \#100) were categorized as major bugs by SonarQube, and the time required to solve the given bug was 15 minutes each. All of these problems were bug-free when their solutions were generated by Amazon CodeWhisperer or ChatGPT. The single bug (Problem ID: \#102) we observed with the solutions of Amazon CodeWhisperer was again a major bug, and the corresponding solutions generated by the other generators were bug-free. The estimated time to solve this bug was five minutes. For ChatGPT, we had a blocker (Problem ID: \#8), and a major bug (Problem ID: \#141). The estimated time to solve the bug in problem \#8 was 15 minutes, and this was 10 minutes for problem \#141. Again, the bugs were unique to ChatGPT among all the generators. 

        Regarding Code Maintainability, we created a list of the code smells encountered in the generated code, shown in Table \ref{tab:list-of-code-smells}. Here we list each smell, accompanied by the frequencies we have encountered for each code generator, the technical debt of the particular smell (estimated time to resolve the issue), and the severity of the smell. As seen in Table \ref{tab:results_Security-Reliability-Maintainability}, 14 problems contained a single, three problems contained two, and 2 problems contained three code smells. The average technical debt for the problems that contained at least one smell was 9.1 minutes and the total estimated time to solve every smell was 172 minutes. For Amazon CodeWhisperer, we have encountered 22 problems where there were single, and 2 problems with two code smells. There were not any instances, which contained more than two code smells. The average technical debt for the problems containing at least one smell was 5.6 minutes, and the total estimated time to solve the smells was 117 minutes. For ChatGPT, there were 13 problems containing a single, and 1 problem containing two code smells. On average, the technical debt of the problems accompanied by at least a single smell was 8.9 minutes. The total estimated time to solve all the smells was 134 minutes.

        \begin{table}
          \caption{List of the Code Smells of the Generated Code}
          \label{tab:list-of-code-smells}
            \begin{tabular}{p{6cm}ccccc}
                {} &  \textbf{Copilot} & \textbf{CodeWhisperer} & \textbf{ChatGPT} & {} & {}\\
                Code Smell &  \multicolumn{3}{c}{Number of Problems} & Technical Debt & Severity\\
                \noalign{\smallskip}\hline \noalign{\smallskip}
                - Rename this variable; it shadows a built-in. & 6 & 2 & 1 & 5 mins & Major \\
                - Remove the unused function parameter. & 1 & 5 & 0 & 5 mins & Major\\
                - Refactor this function to reduce its Cognitive Complexity. & 6 & 2 & 2 & - & Critical\\
                - Merge this if statement with the enclosing one. & 2 & 0 & 1 & 5 mins & Major\\
                - Rename this parameter x to match the regular expression \^{}[\_a-z][a-z0-9\_]*\$. & 2 & 2 & 2 & 2 mins & Minor\\
                - Remove the unused local variable x. & 2 & 0 & 4 & 5 mins & Minor\\
                - Rename function x to match the regular expression \^{}[a-z\_][a-z0-9\_]*\$. & 5 & 5 & 5 & 10 mins & Major\\
                - Specify an exception class to catch or reraise the exception. & 1 & 0 & 0 & 5 mins & Critical\\
                - Extract this nested conditional expression into an independent statement. & 1 & 0 & 0 & 5 mins & Major\\
                - Complete the task associated to this ``TODO" comment. & 0 & 7 & 0 & 0 mins & Info\\
                - Remove commented out code. & 0 & 3 & 0 & 5 mins & Major\\
                - Use concise character class syntax `\symbol{92}d' instead of `[0-9]'. & 0 & 0 & 2 & 5 mins & Minor\\
                - Replace this x call by a y function call. & 0 & 0 & 1 & 2 mins & Critical\\
                \noalign{\smallskip}\hline\noalign{\smallskip}
                \multicolumn{6}{c}{\textbf{Note:} Versions considered for this table: GitHub Copilot - 1.70.8099, Amazon CodeWhisperer - Jan '23, ChatGPT - 9 Jan 2023 Version}
            \end{tabular}

        \end{table}
        
        \subsection{Using only Function Names and Parameters Without Prompt (RQ2)} \label{Results-Using-only-Function-Names-and-Parameters-Without-Prompt}

        \begin{table}
          \caption{Percentage Results of all code generation tools for Original Experiment (ORG), Only Function Name (OFN) and Dummy Function Name (DFN)}
          \label{results_org_ofn_dfn}
          \begin{tabular}{cccccccccc}
            {} &  \multicolumn{3}{c}{\textbf{Copilot v1.70.8099 (New)}} & \multicolumn{3}{c}{\textbf{CodeWhisperer Jan '23 (New)}} & \multicolumn{3}{c}{\textbf{ChatGPT 9 Jan '23 Version}}\\
            \noalign{\smallskip}\hline \noalign{\smallskip}
            {} & ORG & OFN & DFN & ORG & OFN & DFN & ORG & OFN & DFN \\
            \noalign{\smallskip}\hline \noalign{\smallskip}
            \textbf{Valid} & 91.5\% & \textbf{78.0\%} & \textbf{93.9\%} & 90.2\% & \textbf{78.0\%} & 89.6\% & \textbf{93.3\%} & 76.8\% & 92.7\% \\
            \textbf{Correct} & 46.3\% & 20.1\% & 42.1\% & 31.1\% & 14.6\% & 27.4\% & \textbf{65.2\%} & \textbf{22.0\%} & \textbf{61.6\%} \\
            \textbf{Partially Correct} & 23.2\% & 26.8\% & 26.8\% & \textbf{40.2\%} & \textbf{29.9\%} & \textbf{36.6\%} & 22.6\% & 27.4\% & 25.6\% \\
            \textbf{Incorrect} & 30.5\% & 53.1\% & 31.1\% & \textbf{28.7\%} & \textbf{55.5\%} & \textbf{36.0\%} & 12.2\% & 50.6\% & 12.8\% \\
            \noalign{\smallskip}\hline
        \end{tabular}
        \end{table}

        The results we presented up until this point were the outputs of the experiment where we provided the function name, parameters, and the docstring as the inputs to get the generated code. In this part, as we explained in Section \ref{Methodology-Experimental-Setup}, we removed the docstring from each of our problems in the dataset. The results of this experiment are presented in Table \ref{results_org_ofn_dfn}.

        In our original experiment where we used both the function name and the prompt, our code validity score was 91.5\% for GitHub Copilot, 90.2\% for Amazon CodeWhisperer, and 93.3\% for ChatGPT. In our latter experiment, where we only used the function names, our code validity score dropped to 78.0\% for GitHub Copilot, 78.0\% for Amazon CodeWhisperer, and 76.8\% for ChatGPT.
        
        For code correctness, if we compare the results of the two experiments for GitHub Copilot, the rate of correctly generated code dropped from 46.3\% to 20.1\%. The incorrectly generated code percentage increased from 30.5\% to 53.1\%, and the partially correctly generated code percentage increased from 23.2\% to 26.8\%. For Amazon CodeWhisperer, the rate of correctly generated code dropped from 31.1\% to 14.6\%. The incorrectly generated code percentage increased from 28.7\% to 55.5\%, and the partially correctly generated code percentage decreased from 40.2\% to 29.9\%. For ChatGPT, the rate of correctly generated code dropped from 65.2\% to 22.0\%. The incorrectly generated code percentage increased from 12.2\% to 50.6\%, and the partially correctly generated code percentage increased from 22.6\% to 27.4\%.
        
        \subsection{Using Dummy Function Names (RQ3)}\label{Results-Using-Dummy-Function-Names-(RQ3)}

        In this part, as explained in Section \ref{Methodology-Experimental-Setup}, we prompted GitHub Copilot, Amazon CodeWhisperer, and ChatGPT to generate code for the same problems, this time with dummy function names instead of meaningful, and informative function names. We replaced the function names with `foo'. The original and new experiment results are presented in Table \ref{results_org_ofn_dfn}.

        Our code validity score increased to 93.9\% for GitHub Copilot and decreased to 89.6\% for Amazon CodeWhisperer and 92.7\% for ChatGPT.

        For code correctness, if we compare the results of the two experiments for GitHub Copilot, the rate of correctly generated code dropped from 46.3\% to 42.1\%. The incorrectly generated code percentage increased from 30.5\% to 31.1\%, and the partially correctly generated code percentage increased from 23.2\% to 26.8\%. For Amazon CodeWhisperer, the rate of correctly generated code dropped from 31.1\% to 27.4\%. The incorrectly generated code percentage increased from 28.7\% to 36.0\%, and the partially correctly generated code percentage decreased from 40.2\% to 36.6\%. For ChatGPT, the rate of correctly generated code dropped from 65.2\% to 61.6\%. The incorrectly generated code percentage increased from 12.2\% to 12.8\%, and the partially correctly generated code percentage increased from 22.6\% to 25.6\%.

        \subsection{Evaluation of Code Generation Tools Over Time (RQ4)}\label{Results-Evaluation-of-Tools-Over-Time}

        In this part, as explained in Section \ref{Methodology-Evaluation-of-Code-Generation-Tools-Over-Time}, we have evaluated GitHub Copilot and Amazon CodeWhisperer using the newer versions.

\begin{figure*}
                \includegraphics[scale=0.50]{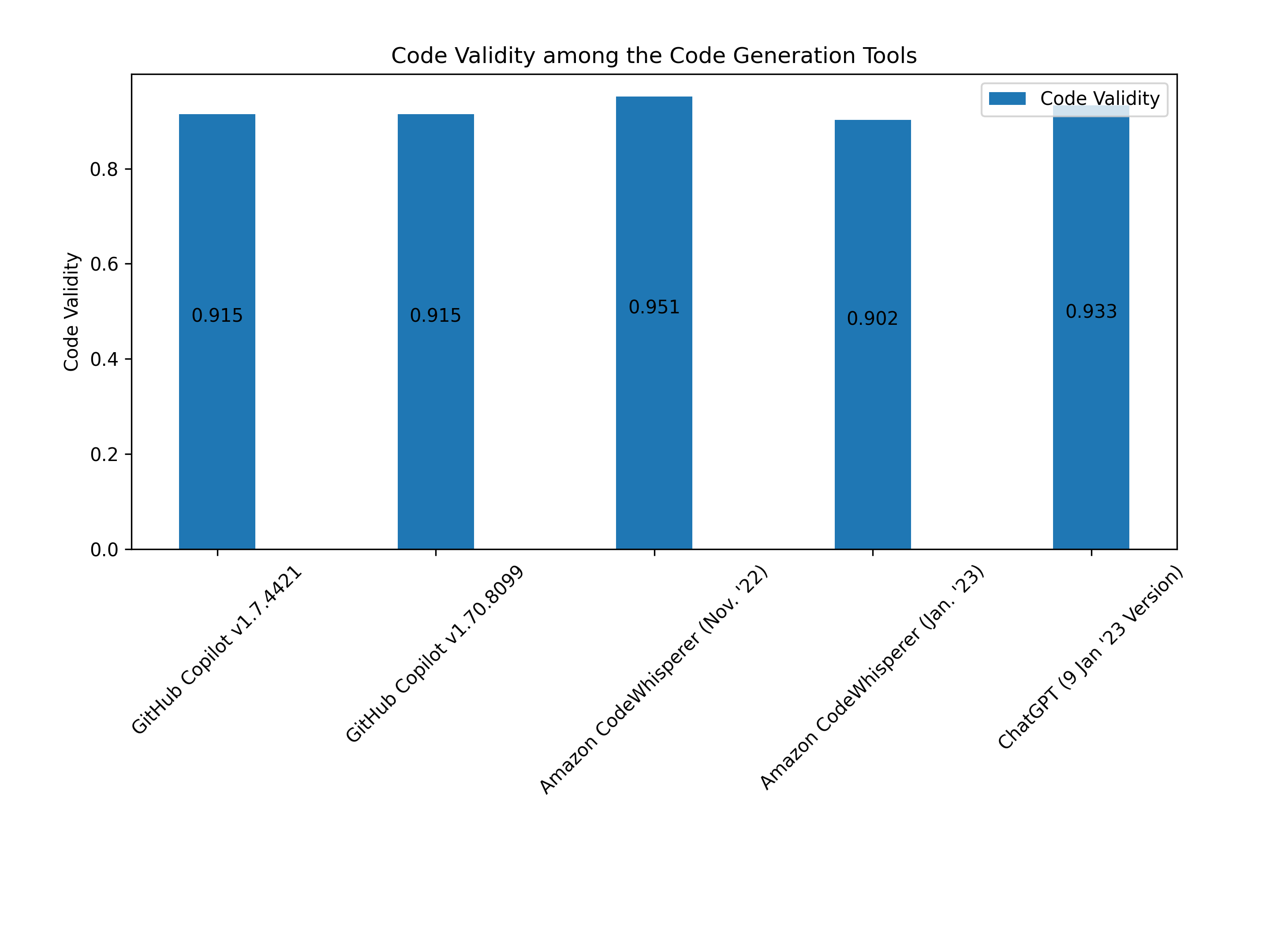}
                \caption{Code Validity Scores of the Code Generation Tools}
                \label{chart_validity_original}
            \end{figure*}

        As shown in Figure \ref{chart_validity_original}, compared to the experiment results where we used older versions of GitHub Copilot and Amazon CodeWhisperer, our code validity score, 91.5\%, did not change for GitHub Copilot and the validity score of Amazon CodeWhisperer, the validity score dropped to 90.2\% (from 95.1\%). 

        For code correctness, if we compare the results of the two experiments for GitHub Copilot, the rate of correctly generated code increased to 46.3\% (from 28.7\%). The incorrectly generated code percentage increased to 30.5\% (from 20.1\%), and the partially correctly generated code percentage decreased to 23.2\% (from 51.2\%). For Amazon CodeWhisperer, the rate of correctly generated code increased to 31.1\% (from 24.4\%). The incorrectly generated code percentage decreased to 28.7\% (from 45.1\%), and the partially correctly generated code percentage increased to 40.2\% (from 30.5\%).

    \section{Discussion}\label{Discussion}
    
        \subsection{Code Validity (RQ1.1)}

            As we discussed, for our 164 problems, GitHub Copilot  was able to generate valid code for 150 of them, yielding a success rate of \begin{math} 91.5\% \end{math}. Amazon CodeWhisperer was able to generate valid code for 148 problems, yielding a success rate of \begin{math} 90.2\% \end{math} and ChatGPT was able to generate valid code for 153 of them, yielding a success rate of \begin{math} 93.3\% \end{math}. 

            The causes of the invalid code generated by GitHub Copilot were operations with incompatible types (Listing \ref{invalid-copilot-137}), syntax errors, and usage of the functions of unimported libraries.
            
            Amazon CodeWhisperer had the following causes preventing a particular code from being valid: usage of the functions of unimported libraries (Listing \ref{invalid-whisperer-162}), improper list indexing, operations with incompatible types, searching for values that are not in a particular list (Listing \ref{invalid-whisperer-70}), incorrect usage of the assert statements, syntax errors, and stack overflow errors.

            Lastly, ChatGPT had the following causes for invalid code: improper list and string indexing, syntax errors (Listing \ref{invalid-chatgpt-8}), operations with incompatible types (Listing \ref{invalid-chatgpt-95}), and the usage of the functions of unimported libraries.

            From these results, we can see that there were many common issues among the code generation tools that were the causes of invalid code. While the frequencies of these issues were not unique among the tools, the small number of invalid codes should refrain us to make a generalization of any issue to correspond to a particular tool more than some other one. We argue that the occurrence of similar issues among the tools, also the similar rates of success of the code generation tools suggest that they are practically similar to each other in terms of being able to successfully generate valid code. The approximation appears to be that the code generation tools are able to generate valid code 9 out of 10 times.  In the generated code, some issues like syntax errors are more visible to the programmer, than for example operations with incompatible types. When the latter occurs in a given code, it is less unlikely that the programmer notices this issue since in some instances the code can run without any errors for a given input; however, fail for another one. Therefore we want to highlight this particular vulnerability of the generated code by the tools.
            
            \medskip
            \fbox{\begin{minipage}{33em}
                All code generation tools are capable of generating valid code 9 out of 10 times with mostly similar types of issues. The practitioners should expect that for 10\% of the time the generated code by the code generation tools would be invalid. Moreover, they should test their code thoroughly to catch all possible cases that may cause the generated code to be invalid.
            \end{minipage}}

            \begin{figure}[h]
\begin{lstlisting}[caption={Invalid Code Generation by GitHub Copilot (ID: 137)}, label={invalid-copilot-137}, numbers=none]
def compare_one(a, b):
    """
    Create a function that takes integers, floats, or strings representing real numbers, and returns the larger variable in its given variable type. Return None if the values are equal. Note: If a real number is represented as a string, the floating point might be . or ,

    compare_one(1, 2.5) --> 2.5
    compare_one(1, "2,3") --> "2,3"
    compare_one("5,1", "6") --> "6"
    compare_one("1", 1) --> None
    """
    if a == b:
        return None
    return max(a, b)
\end{lstlisting}
            \end{figure}
            

            \begin{figure}[h]
\begin{lstlisting}[caption={Invalid Code Generation by Amazon CodeWhisperer (ID: 162)}, label={invalid-whisperer-162}, numbers=none]
def string_to_md5(text):
    """
    Given a string 'text', return its md5 hash equivalent string.
    If 'text' is an empty string, return None.

    >>> string_to_md5('Hello world') == '3e25960a79dbc69b674cd4ec67a72c62'
    """
    if text == '':
        return None
    else:
        return hashlib.md5(text).hexdigest()
\end{lstlisting}
            \end{figure}



            \begin{figure}[h]
\begin{lstlisting}[caption={Invalid Code Generation by Amazon CodeWhisperer (ID: 70)}, label={invalid-whisperer-70}, numbers=none]
def strange_sort_list(lst):
    '''
    Given list of integers, return list in strange order. Strange sorting, is when you start with the minimum value, then maximum of the remaining integers, then minimum and so on.

    Examples:
    strange_sort_list([1, 2, 3, 4]) == [1, 4, 2, 3]
    strange_sort_list([5, 5, 5, 5]) == [5, 5, 5, 5]
    strange_sort_list([]) == []
    '''
    if not lst:
        return lst
    min_val = min(lst)
    max_val = max(lst)
    lst.remove(min_val)
    lst.remove(max_val)
    return [min_val] + [max_val] + strange_sort_list(lst)
\end{lstlisting}
            \end{figure}
            



            \begin{figure}[h]
\begin{lstlisting}[caption={Invalid Code Generation by ChatGPT (ID: 8)}, label={invalid-chatgpt-8}, numbers=none]
from typing import List, Tuple

def sum_product(numbers: List[int]) -> Tuple[int, int]:
    """ For a given list of integers, return a tuple consisting of a sum and a product of all the integers in a list. Empty sum should be equal to 0 and empty product should be equal to 1.
    """
    if not numbers:
        return (0, 1)
    else:
        return (sum(numbers), 1)

for i in numbers:
    result *= i

return (result, sum_result)
\end{lstlisting}
            \end{figure}

            \begin{figure}[h]
\begin{lstlisting}[caption={Invalid Code Generation by ChatGPT (ID: 95)}, label={invalid-chatgpt-95}, numbers=none]
def check_dict_case(dict):
    if not dict:
        return False
    lower = all(key.islower() for key in dict)
    upper = all(key.isupper() for key in dict)
    return lower or upper
\end{lstlisting}
            \end{figure}


        \subsection{Code Correctness (RQ1.2)}

        \begin{figure*}
            \includegraphics[scale=0.6]{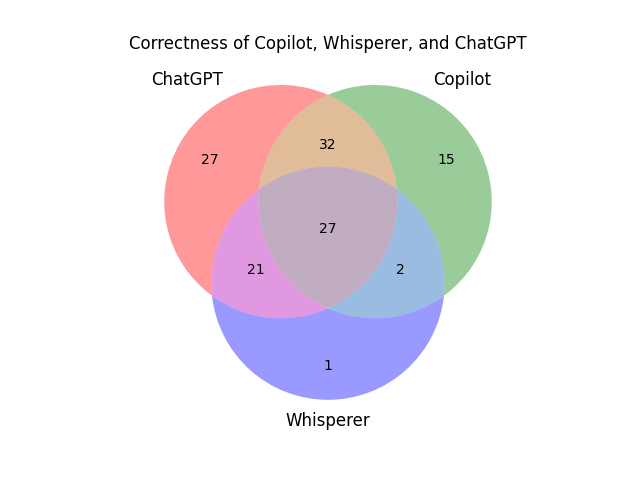}
            \caption{Distribution of Correctly Generated Samples among the Code Generation Tools}
            \label{code_correctness_distribution}
        \end{figure*}

        In our previous study, before we obtained our results, we hypothesized that code generation tools could either understand the user intent or not \citep{yetistiren2022assessing}. This characteristic would result in a correct or incorrect code for a given problem.

        However, like our previous study, here as well; in contrast to a binary scenario for code correctness where the generated solution is either correct or incorrect, code generation tools also generated partially correct solutions. As shown in Figure \ref{wheel_chart_copilot_a}, for \begin{math} 23.2\% \end{math} of the problems, GitHub Copilot generated partially correct code and in Figure \ref{wheel_chart_copilot_b}, it can be seen that \begin{math} 60.4\% \end{math} of the partially correct generations have correctness score above \begin{math} 50.0\% \end{math}.

        As it can be observed in Figure \ref{wheel_chart_whisperer_a}, for \begin{math} 40.2\% \end{math} of the problems, Amazon CodeWhisperer generated partially correct code and in Figure \ref{wheel_chart_whisperer_b}, it can be seen that \begin{math} 53.1\% \end{math} of the partially correct generations have correctness score above \begin{math} 50.0\% \end{math}.

        As it can be observed in Figure \ref{wheel_chart_chatgpt_a}, for \begin{math} 22.6\% \end{math} of the problems, ChatGPT generated partially correct code and in Figure \ref{wheel_chart_chatgpt_b}, it can be seen that \begin{math} 63.2\% \end{math} of the partially correct generations have correctness score above \begin{math} 50.0\% \end{math}.

        Therefore, we argue that not only the entirely correct solutions should be considered a success, but the partially correct solutions should also be taken into account. This is the case because usually in regular programming practices, it is seen that the first iteration made on the written code is not correct. Still, over the next iterations, the code becomes correct. Therefore, we argue that GitHub Copilot, Amazon CodeWhisperer, and ChatGPT are considerably successful in understanding user intent and generating the correct code. Our standpoint is further elaborated on with an example in the following paragraphs.

        For example, in Listing \ref{partially-correct-chatgpt-10}, we give an example of partially correctly generated code by ChatGPT. The generated code by ChatGPT defines the bounds and the increment of the for loop with (1). The code correctness score for this approach is 20\%. However, if we change the bounds and the increment of the same for loop as shown in (2), the code correctness score increases to 100\%. In Listing \ref{partially-correct-whisperer-54}, we present the code generated for problem \#54. The return statement marked with (1) is the Amazon CodeWhisperer's generation. This solution to the problem has a 57\% code correctness score. The return statement marked with (2) is the correct approach for the solution to the problem. Such examples show that, even with minor additional input from the programmer, the code correctness scores can be improved. We have only provided two examples for Amazon CodeWhisperer and ChatGPT; however, the examples for each code generation tool can be diversified including GitHub Copilot. 

        \begin{figure}[h]
\begin{lstlisting}[caption={Partially Correct Code Generation by ChatGPT (ID: 10)}, label={partially-correct-chatgpt-10}, numbers=none]
def is_palindrome(string: str) -> bool:
    """ Test if given string is a palindrome """
    return string == string[::-1]

def make_palindrome(string: str) -> str:
    """ Find the shortest palindrome that begins with a supplied string.
    Algorithm idea is simple:
    - Find the longest postfix of supplied string that is a palindrome.
    - Append to the end of the string reverse of a string prefix that comes before the palindromic suffix.
    """
    if not string:
        return ''
(1) for i in range(len(string), 0, -1): 
(2) for i in range(0, len(string), 1): 
        if is_palindrome(string[i:]):
            return string + string[:i][::-1]
\end{lstlisting}
        \end{figure}

        \begin{figure}[h]
\begin{lstlisting}[caption={Partially Correct Code Generation by Amazon CodeWhisperer (ID: 54)}, label={partially-correct-whisperer-54}, numbers=none]
def same_chars(s0: str, s1: str):
    """
    Check if two words have the same characters.
    >>> same_chars('eabcdzzzz', 'dddzzzzzzzddeddabc')
    True
    >>> same_chars('abcd', 'dddddddabc')
    True
    >>> same_chars('dddddddabc', 'abcd')
    True
    >>> same_chars('eabcd', 'dddddddabc')
    False
    >>> same_chars('abcd', 'dddddddabce')
    False
    >>> same_chars('eabcdzzzz', 'dddzzzzzzzddddabc')
    False
    """
(1) return sorted(s0) == sorted(s1)
(2) return set(s0) == set(s1)
\end{lstlisting}
        \end{figure}

        In Listing \ref{easy-and-difficult}, we exemplify our other finding about the code generation tools we have evaluated. The solutions to two functions (``common" and ``minPath") are generated by GitHub Copilot. As it can be seen the prompt for the ``common" function is more straightforward, whereas the prompt for the ``minPath" function is more complicated. Such characteristics of these prompts are reflected in the code generation performance of GitHub Copilot; for the first function, a correct solution is generated, and for the latter one GitHub Copilot failed to generate any algorithm. Similar to our previous finding, this tendency can also be generalized to all of the code generation tools that we evaluated.

        \begin{figure}[h]
\begin{lstlisting}[caption={Code Generation by GitHub Copilot for a Problem with Easier, and More Complicated Prompt}, label={easy-and-difficult}, numbers=none]
def common(l1: list, l2: list):
    """Return sorted unique common elements for two lists.
    >>> common([1, 4, 3, 34, 653, 2, 5], [5, 7, 1, 5, 9, 653, 121])
    [1, 5, 653]
    >>> common([5, 3, 2, 8], [3, 2])
    [2, 3]

    """
    result = []
    for i in l1:
        if i in l2 and i not in result:
            result.append(i)
    return sorted(result)

######################################################

def minPath(grid, k):
    """
    Given a grid with N rows and N columns (N >= 2) and a positive integer k, each cell of the grid contains a value. Every integer in the range [1, N * N] inclusive appears exactly once on the cells of the grid.

    You have to find the minimum path of length k in the grid. You can start from any cell, and in each step you can move to any of the neighbor cells, in other words, you can go to cells which share an edge with you current cell. Please note that a path of length k means visiting exactly k cells (not necessarily distinct). You CANNOT go off the grid. A path A (of length k) is considered less than a path B (of length k) if after making the ordered lists of the values on the cells that A and B go through (let's call them lst_A and lst_B), lst_A is lexicographically less than lst_B, in other words, there exist an integer index i (1 <= i <= k) such that lst_A[i] < lst_B[i] and for any j (1 <= j < i) we have lst_A[j] = lst_B[j]. It is guaranteed that the answer is unique. Return an ordered list of the values on the cells that the minimum path go through.

    Examples:

        Input: grid = [ [1,2,3], [4,5,6], [7,8,9]], k = 3
        Output: [1, 2, 1]

        Input: grid = [ [5,9,3], [4,1,6], [7,8,2]], k = 1
        Output: [1]
    """
    pass 
\end{lstlisting}
        \end{figure}

        To discuss the comparative code generation success of GitHub Copilot, Amazon CodeWhisperer, and ChatGPT, we created Figure \ref{code_correctness_distribution}. The Venn diagram in this figure shows us the correct code generation capabilities of the code generation tools more in-depth, in comparison to the mere percentage values. With 36 unique problems, ChatGPT managed to generate the correct solution for more problems than GitHub Copilot and Amazon CodeWhisperer. This is followed by GitHub Copilot with 15 problems, and two problems with Amazon CodeWhisperer. Moreover, GitHub Copilot and ChatGPT generated correct solutions to 29 problems where Amazon CodeWhisperer failed; this number was 19 problems for the union of ChatGPT and Amazon CodeWhisperer. Our findings from this evaluation can also be supported by the percentage values. These were 46.3\% Code Correctness and 59.85\% Average Code Correctness for GitHub Copilot; 31.1\% Code Correctness and 51.95\% Average Code Correctness for Amazon CodeWhisperer; and 65.2\% Code Correctness and 78.1\% Average Code Correctness for ChatGPT.

        \medskip
            \fbox{\begin{minipage}{33em}
                For better Code Correctness scores, continuous input from the practitioners is needed for all of the code generation tools we evaluated. Additionally, we have found that the generated solutions for longer and more complex prompts for the functions yielded lower Code Correctness scores, in contrast to the functions that had simpler instructions contained in the prompt. For individual Code Correctness performances of the code generation tools, we have found that ChatGPT was the most successful and Amazon CodeWhisperer was the least successful tool. Practitioners that will potentially employ these tools should await similar results for the correctness of their code generated by these code generation tools.
            \end{minipage}}

        \subsection{Code Security \& Code Reliability \& Code Maintainability (RQ1.3 \& RQ1.4 \& RQ1.5)}\label{Discussion-Code-Security-Reliability-Maintainability}

\begin{figure*}
            \includegraphics[scale=0.55]{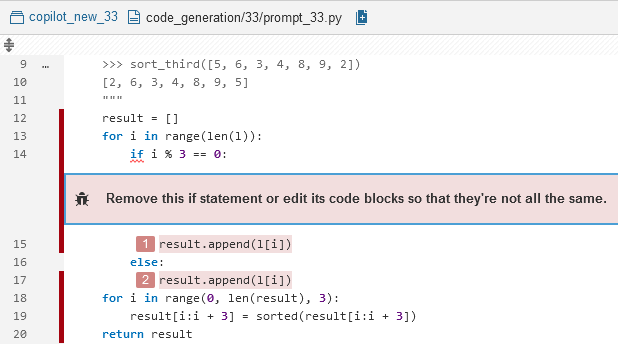}
            \caption{GitHub Copilot v1.70.8099 Bug in Problem \#33}
            \label{fig:copilot_bug_33}
        \end{figure*}

\begin{figure*}
            \includegraphics[scale=0.55]{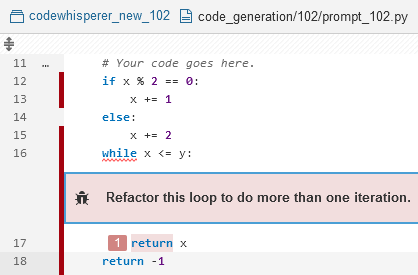}
            \caption{Amazon CodeWhisperer Jan '23 Bug in Problem \#102}
            \label{fig:whisperer_bug_102}
        \end{figure*}

\begin{figure*}
            \includegraphics[scale=0.55]{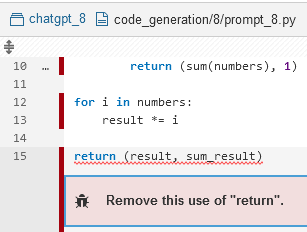}
            \caption{ChatGPT 9 Jan '23 version Bug in Problem \#8}
            \label{fig:chatgpt_bug_8}
        \end{figure*}

\begin{figure*}
            \includegraphics[scale=0.55]{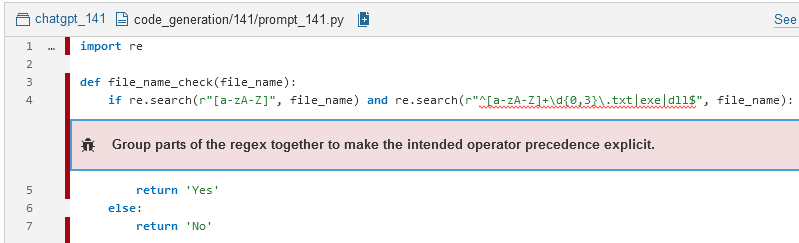}
            \caption{ChatGPT 9 Jan '23 version Bug in Problem \#141}
            \label{fig:chatgpt_bug_141}
        \end{figure*}

\begin{figure*}
            \includegraphics[scale=0.7]{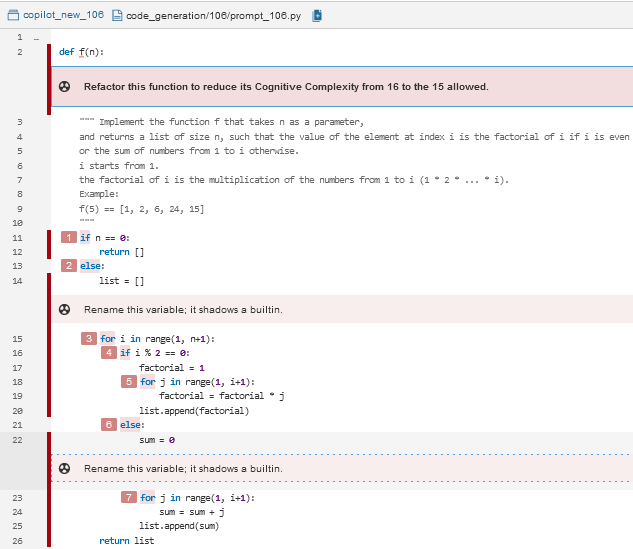}
            \caption{GitHub Copilot v1.70.8099 Smells (3) in Problem \#106}
            \label{fig:copilot_smell_106}
        \end{figure*}

\begin{figure*}
            \includegraphics[scale=0.6]{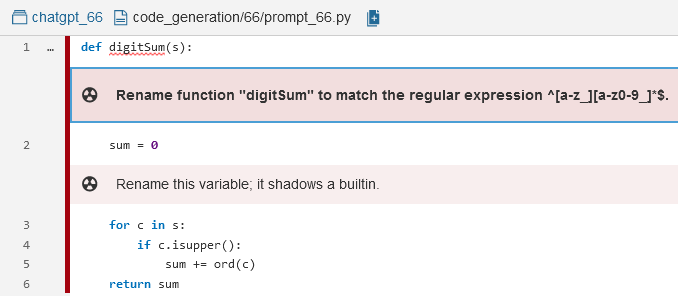}
            \caption{ChatGPT 9 Jan '23 version Smells (2) in Problem \#66}
            \label{fig:chatgpt_smell_66}
        \end{figure*}

        \begin{itemize}[leftmargin=0cm]

            \item[] \textbf{Code Security:} As explained in Section \ref{Results-Code-Security-Reliability-Maintainability}, we have seen no difference between the generators in terms of Code Security. Moreover, the security rating of all of the problems for each generator had the maximum rating, therefore per our results and our benchmark dataset, we can say that the generators are equally successful in terms of generating secure code. \\
            
            \item[] \textbf{Code Reliability:} Our Code Reliability results were represented by the number of bugs observed in each sample, the severity of these particular bugs, and the estimated time to solve them. In Figures \ref{fig:copilot_bug_33} one of the bugs can be observed. The cause of this bug is the inconsistent usage of the `if' statement, that the same expression is written under different conditions. The other bugs (Problem \#37 and \#100) we have found for GitHub Copilot have the same cause as the previous problem. In Figure \ref{fig:whisperer_bug_102}, the only bug we found by SonarQube for Amazon CodeWhisperer is shown. The cause of this bug is the single iteration of a `while' loop. The bugs for ChatGPT can be seen in Figures \ref{fig:chatgpt_bug_8} and \ref{fig:chatgpt_bug_141}. One of these bugs was caused by the wrong indentation usage for the `return' statement and the other one was caused by the regular expression, that the \^ character has higher precedence and the expression would be anchored upon. \\

            The vulnerabilities therefore should be taken into account by the potential users of the code generation tools, and they might reflect some of the weaknesses of the code generators and be showing one of the possible directions of improvement. As we have argued, the generators have unique bugs, and the bugs do not correspond to the generated solution for the same problem by other generators. Therefore, we cannot prove the superiority of a given generator to another regarding its reliable code generation capabilities. From another perspective, regarding the estimated time to eliminate the bugs, we have seen that the most successful candidate was Amazon CodeWhisperer with 5 minutes on average, and the least successful candidate was GitHub Copilot, with 15 minutes. The average estimated time to eliminate the bugs contained in the code generated by ChatGPT was 12.5 minutes. While from this perspective, there appears to be a ranking among the code generation tools, we believe that considering our previous point too, a conclusion regarding the bug-free code generation capabilities of these tools should not be made solely relying on the estimated time to eliminate the bugs.\\

            \item[] \textbf{Code Maintainability:} During our evaluation, we have seen some code smells among all the generators, varying in severity and resulting in a considerable amount of Technical Debt. The exact results were shown in Table \ref{tab:results_Security-Reliability-Maintainability} and explained in Section \ref{Results-Code-Security-Reliability-Maintainability}. The three most common issues were, improper naming of the function or variable, and high cognitive complexities.\\

            In Figure \ref{fig:copilot_smell_106}, the instances of the code with high cognitive complexity and improper naming of variables can be seen. This particular example was generated by GitHub Copilot; however, the smells in this code were also seen in some of the code generated by Amazon CodeWhisperer and ChatGPT. Figure \ref{fig:chatgpt_smell_66} shows the other most common type of smell, which is the improper naming of the function. There is also an additional smell, which is the improper naming of a variable, which is named after a reserved word in Python. The naming of the function is found to be improper by SonarQube since it adopted the camel case approach, which should have been the Snake Case approach for Python. \\

            In general, we have seen that our findings for Code Maintainability of the generated code by GitHub Copilot, Amazon CodeWhisperer, and ChatGPT, where we have seen some shortcomings were the most noteworthy ones. For Code Security and Code Reliability metrics, we did not find significant results, which we could generalize to all the code generators, or show a given generator was performing better than the others. However, we could generalize the Code Maintainability results; they allowed us to list all of the code smells we observed using our benchmark dataset. Most importantly, apart from some smells, we could see them in the code generated by all of the generators.       
        \end{itemize}

        \medskip
        \fbox{\begin{minipage}{33em}
           Code that contains some bugs should be excepted by the practitioners of GitHub Copilot, Amazon CodeWhisperer, and ChatGPT. However, per our results, they are not as common as the code smells, which we have observed for all code generators. Practicioners should note that if their code contains some smells, the average time to solve them is 9.1 minutes for GitHub Copilot, 5.6 minutes for Amazon CodeWhisperer, and 8.9 minutes for ChatGPT. In terms of Code Security, our results showed that the practitioners should await to get secure code from the generators.
        \end{minipage}}

        \subsection{Using only Function Names and Parameters Without Prompt (RQ2)} \label{disc_UoFNaPWP}

        \begin{figure*}
            \includegraphics[scale=0.70]{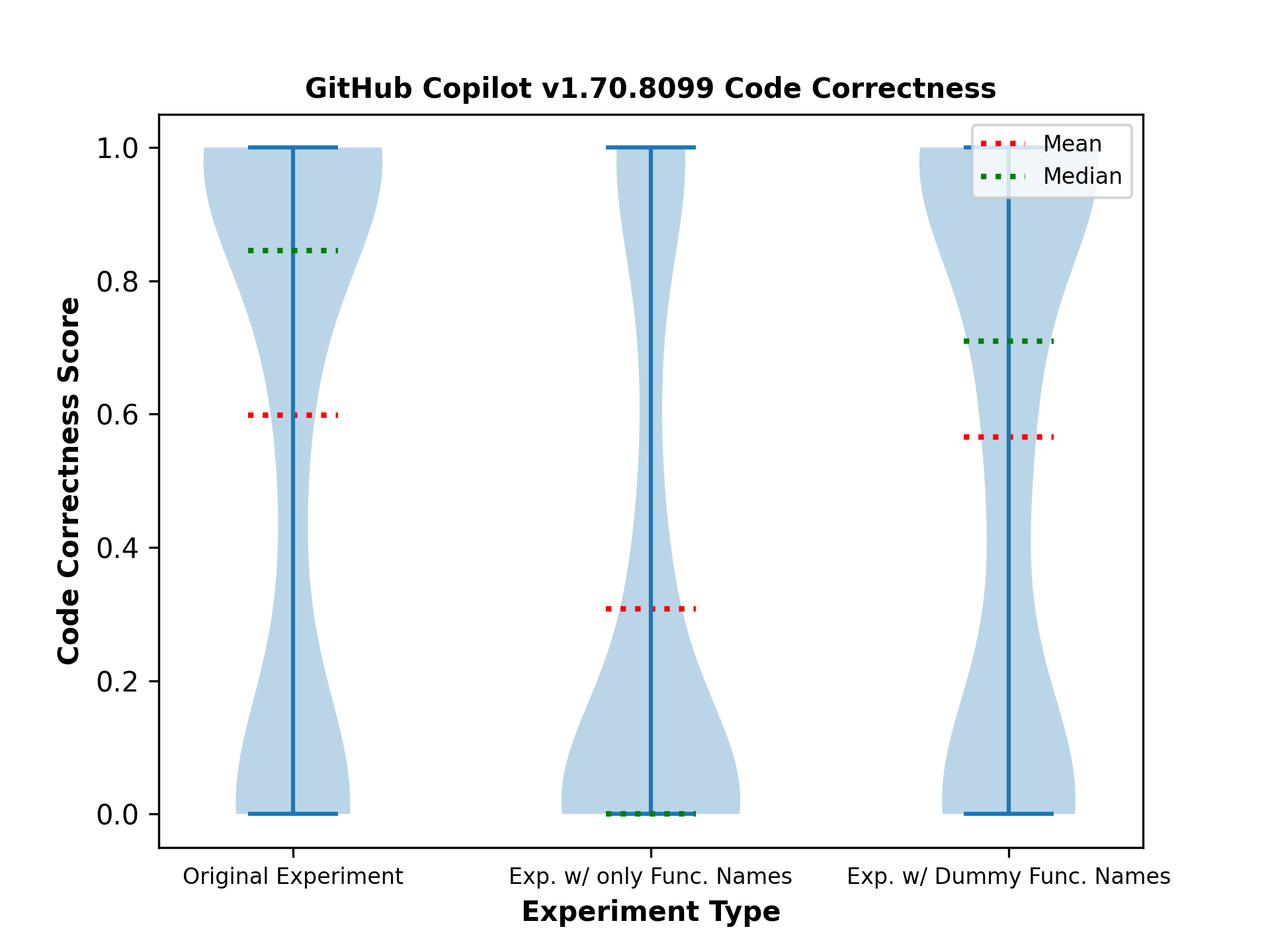}
            \caption{Code Correctness Score Distribution of the Problems for Different Experiments - GitHub Copilot v1.70.8099}
            \label{fig:violin_copilot}
        \end{figure*}

        \begin{figure*}
            \includegraphics[scale=0.70]{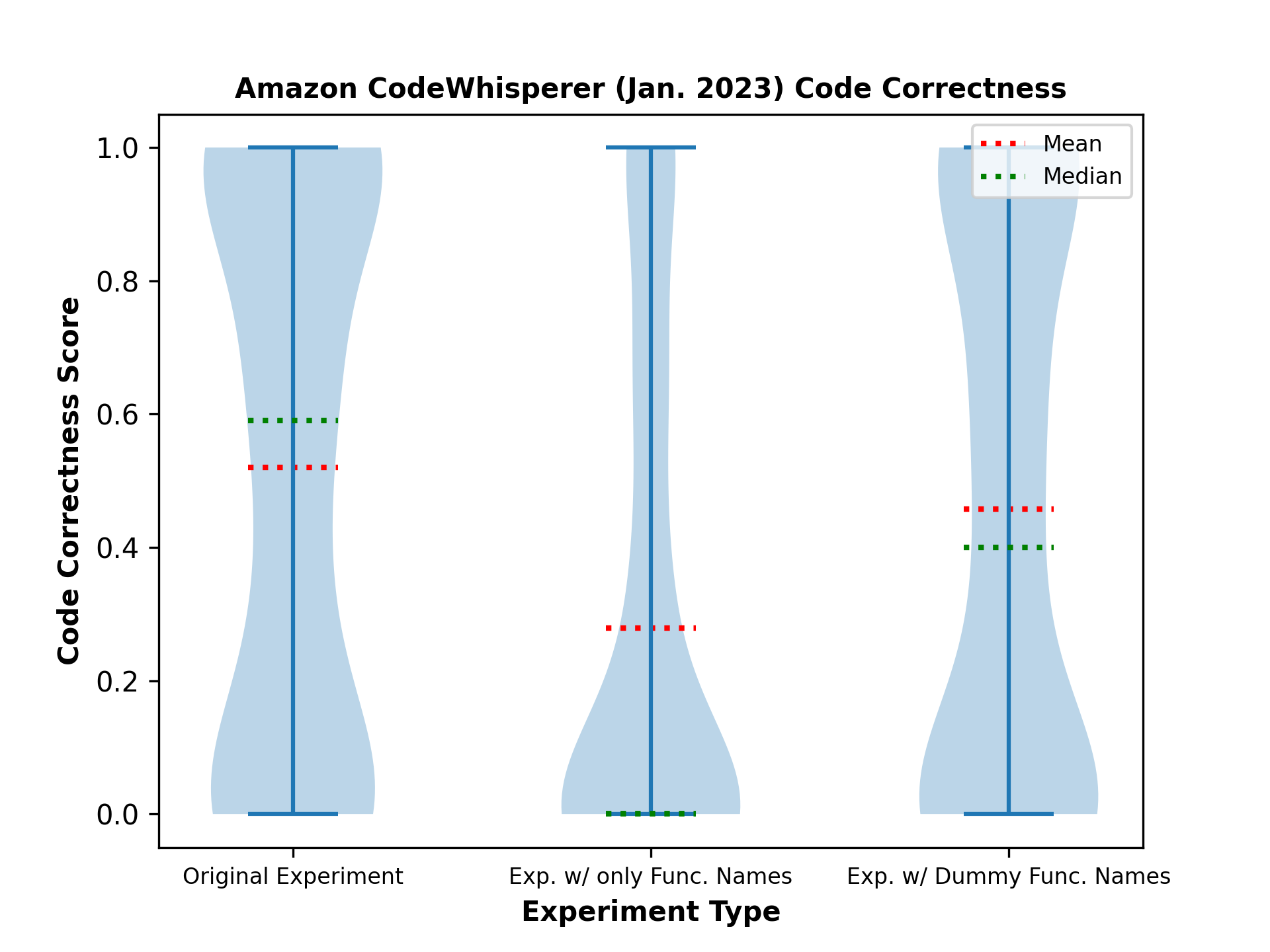}
            \caption{Code Correctness Score Distribution of the Problems for Different Experiments - Amazon CodeWhisperer Jan '23}
            \label{fig:violin_whisperer}
        \end{figure*}
        
        \begin{figure*}
            \includegraphics[scale=0.70]{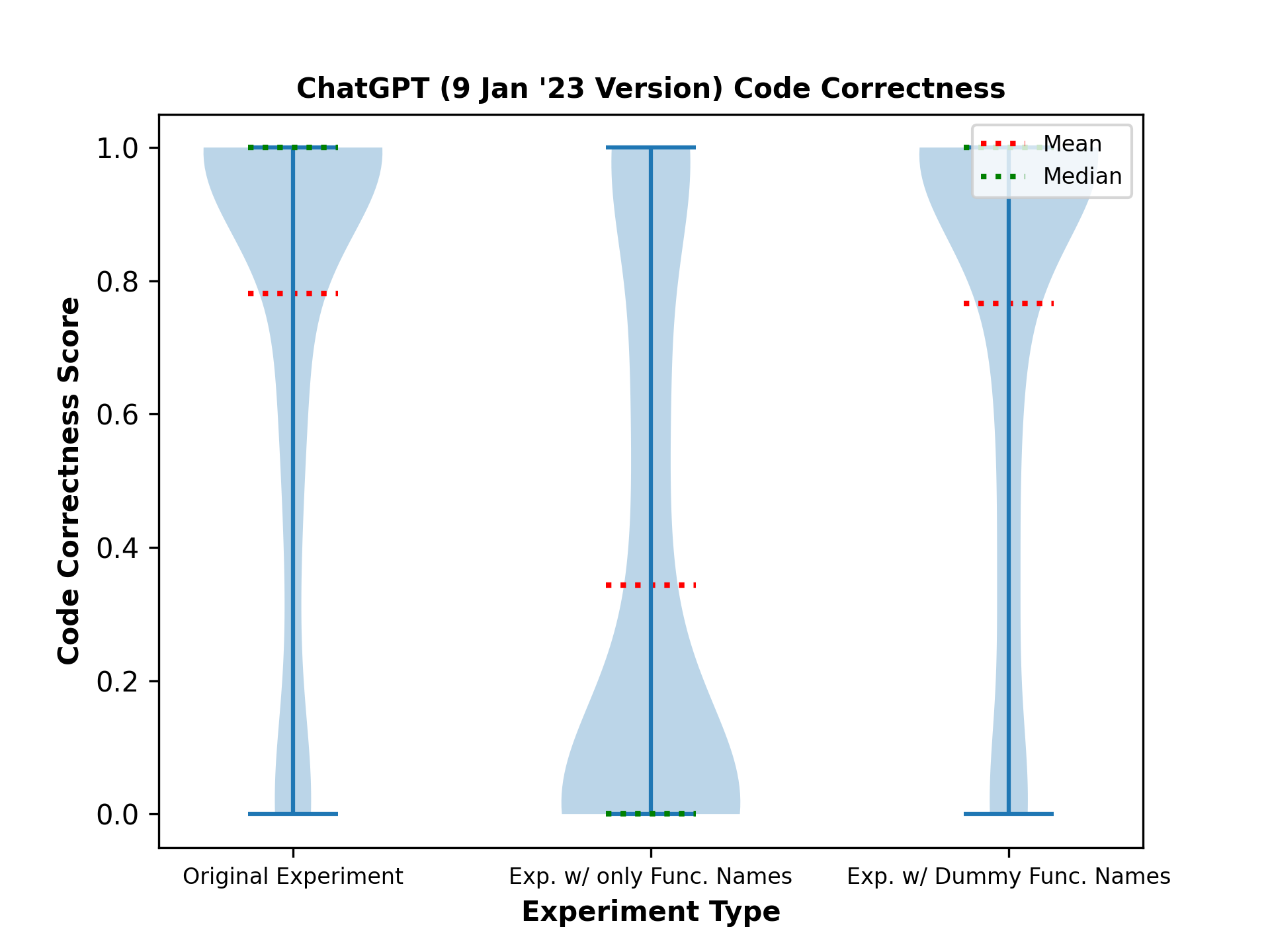}
            \caption{Code Correctness Score Distribution of the Problems for Different Experiments - ChatGPT 9 Jan '23 Version}
            \label{fig:violin_chatgpt}
        \end{figure*}

        According to the results presented in Section \ref{Results-Using-only-Function-Names-and-Parameters-Without-Prompt}, we observed a significant drop in both the code validity and code correctness metrics. For example, the code validity dropped by 13.5\% for GitHub Copilot, 12.2\% for Amazon CodeWhisperer, and 16.5\% for ChatGPT after the docstrings were removed from the problems. Similarly, the code correctness score dropped by 26.2\%, 16.5\%, and 43.2\% respectively. These results reflected a general performance drop affecting the validity and the correctness of the code. From this, we argue that the code generation performance of code generation tools correlates with the input explanation. In Figures \ref{fig:violin_copilot},  \ref{fig:violin_whisperer}, and \ref{fig:violin_chatgpt} the distribution for the code correctness scores of the code generation tools are visualized. Through these distributions, it is trivial to discern that the correct code generation capabilities of the tools tend to be affected negatively. This is demonstrated by the cumulation of problems on the lower half part of the distributions, and the dropped mean and median values. From this, we argue that the lack of a proper explanation of the problems yields lower validity and correctness scores. Therefore, in practical usage, practitioners should pay attention to providing instructions for the code they tend to write to the tools.

        There were some cases, where we did not see any decrease in correctness or validity scores. For GitHub Copilot, such problems constituted 82.3\% of the dataset for code validity and 45.1\% for code correctness. For Amazon CodeWhisperer,  in 84.1\% of the problems for code validity and 63.4\% for code correctness, we did not observe a decrease. For ChatGPT, the scores did not decrease for 78.7\% of the dataset for code validity and 51.2\% for code correctness. We have seen such cases mostly for problems that include substring search, value manipulations in an array, and character comparison. Additionally, the names of such functions, accompanied by parameter names were self-explanatory, which means that GitHub Copilot could still make interpretations about the function without requiring more details.

        For the cases where the code correctness and validity scores dropped, we observed that these problems were more complicated. When we examined where the success rate of code generation tools dropped, we observed cases where the function name and the parameters alone failed to give details. This means that the name and parameters alone are not informative enough to give details about such functions. For example, in one case, a function called “will\_it\_fly” only has two parameters called `q' and `w'. Amazon CodeWhisperer and ChatGPT generated the correct code in our original experiment where we used the function name and prompt but after removing the function explanation from the input, they were not able to generate valid code. To be more precise, ChatGPT could not generate any code at all. The purpose of the function was to check if `q' was a palindromic list and if the sum of the elements in the list was less than `w'. The generations of Amazon CodeWhisperer with and without prompt can be seen in Listing \ref{prompt-72-whisperer}.

        \begin{figure}[h]
\begin{lstlisting}[caption={Generated Code of Amazon CodeWhisperer for the Example Problem (ID: 72)}, label={prompt-72-whisperer},  ]
# Code generation of Amazon CodeWhisperer with prompt
def will_it_fly(q,w):
    '''
    Write a function that returns True if the object q will fly, and False otherwise. The object q will fly if it's balanced (it is a palindromic list) and the sum of its elements is less than or equal the maximum possible weight w.

    Example:
    will_it_fly([1, 2], 5) --> False 
    # 1+2 is less than the maximum possible weight, but it's unbalanced.

    will_it_fly([3, 2, 3], 1) --> False
    # it's balanced, but 3+2+3 is more than the maximum possible weight.

    will_it_fly([3, 2, 3], 9) --> True
    # 3+2+3 is less than the maximum possible weight, and it's balanced.

    will_it_fly([3], 5) --> True
    # 3 is less than the maximum possible weight, and it's balanced.
    '''
    return sum(q) <= w and q == q[::-1]


# Code generation of Amazon CodeWhisperer without prompt
def will_it_fly(q, w):
    if q == 0:
        return "NO"
    elif q == 1:
        return "YES"
    else:
        return will_it_fly(q-w, w)
\end{lstlisting}
        \end{figure}

        \medskip
        \fbox{\begin{minipage}{33em}
         When using code generation tools, it is crucial to provide clear and accurate problem descriptions to obtain valid and correct code. Whenever possible, programmers should include a comprehensive explanation of the problem, along with sample unit tests in the form of docstrings, comments, or other forms of documentation during the solution generation process.
        \end{minipage}}

        \subsection{Using Dummy Function Names (RQ3)}

        According to the results in Section \ref{Results-Using-Dummy-Function-Names-(RQ3)}, we did not observe a dramatic change in the code validity scores, while there was a drop in the code correctness scores. For example, the code validity increased by 2.4\% for GitHub Copilot and dropped by 0.6\% for Amazon CodeWhisperer and ChatGPT after the function names became `foo'. The code correctness score dropped by 4.2\%, 3.7\%, and 3.6\%, respectively. These results reflected a general performance affecting the validity did not change much but it dropped for the correctness of the code. From this, we argue that the code generation performance of code generation tools correlates with the input explanation. Figures \ref{fig:violin_copilot},  \ref{fig:violin_whisperer}, and \ref{fig:violin_chatgpt} show our point visually as well. However, as one may notice, the change in the distributions, in other words, the cumulation of the problems on the lower half, and the drop of the mean and median values were not as radical as we observed them for the experiments with only function names. Hence, we argue that while providing the proper function names is crucial itself, the effect of not providing a proper explanation is more drastic.

        \begin{figure}[h]
\begin{lstlisting}[caption={Generated Codes for the Example Problem (ID: 79)}, label={prompt-79-codeWhisperer},  ]
def decimal_to_binary(decimal):
    """You will be given a number in decimal form and your task is to convert it to binary format. The function should return a string, with each character representing a binary number. Each character in the string will be '0' or '1'. There will be an extra couple of characters 'db' at the beginning and at the end of the string. The extra characters are there to help with the format.
    Examples:
    decimal_to_binary(15)   # returns "db1111db"
    decimal_to_binary(32)   # returns "db100000db"
    """

    #Result of Amazon CodeWhisperer in the original experiment
    return 'db' + bin(decimal)[2:] + 'db'

    #Result of Amazon CodeWhisperer in the dummy experiment
    return sum([1 for i in num if i in "2357BCDE"])    
\end{lstlisting}
        \end{figure}

        According to the majority of our results, the absence of clear and descriptive function names is one of the primary causes of the decline in code correctness and average code correctness. To illustrate, as shown in Listing \ref{prompt-79-codeWhisperer}, in problem \#79 it is asked to convert the given decimal number to binary and add `db' to both the start and end of the string and return it. In the original experiment, its name was `decimal\_to\_binary', which is a highly descriptive name considering the purpose of the function. Hence, Amazon CodeWhisperer resulted in 100\% correctness in the original setup. When we eliminated the function name, we observed that the solution to the problem converted to an invalid one. On the other hand, GitHub Copilot and ChatGPT successfully solved this problem in both setups. When we considered problem \#1, it asked to separate the parenthesis groups given in one string. The method should find the nested groups and separate them. The function's name was `separate\_paren\_groups' in the original setup and it can be seen in Listing \ref{prompt-01-chatgpt}. While ChatGPT had 100\% correctness in the original setup, it resulted in 0\% correctness in the dummy function name setup, which suggests the importance of the proper function name for the code generation tools.

        \begin{figure}[h]
\begin{lstlisting}[caption={Generated Codes for the Example Problem (ID: 01)}, label={prompt-01-chatgpt},  ]
from typing import List
def separate_paren_groups(paren_string: str) -> List[str]:
    """ Input to this function is a string containing multiple groups of nested parentheses. Your goal is to separate those group into separate strings and return the list of those. Separate groups are balanced (each open brace is properly closed) and not nested within each other Ignore any spaces in the input string.
    """

    
    #Result of ChatGPT in the original experiment
    groups = []
    current_group = []
    balance = 0
    for ch in paren_string:
        if ch == ' ':
            continue
        elif ch == '(':
            balance += 1
            current_group.append(ch)
        elif ch == ')':
            balance -= 1
            current_group.append(ch)
            if balance == 0:
                groups.append(''.join(current_group))
                current_group = []
    return groups

    #Result of ChatGPT in the dummy experiment
    paren_string = paren_string.replace(" ", "")
    groups = []
    stack = []
    start = 0
    for i, char in enumerate(paren_string):
        if char == "(":
            stack.append(i)
        elif char == ")":
            if len(stack) > 0:
                start = stack.pop()
                groups.append(paren_string[start:i+1])
    return groups
\end{lstlisting}
        \end{figure}

        Considering the results given in Table \ref{results_org_ofn_dfn}, even though the code validity scores are close, the code correctness scores decreased. It can be concluded that changing the meaningful function names with dummy function names reduced the performance of Copilot for most of the problems. It can be stated that generally changing meaningful function names to dummy function names affects the performance of GitHub Copilot negatively compared to the original experiment where we use both meaningful function names and prompts.

        \medskip
        \fbox{\begin{minipage}{33em}
         Selecting a meaningful name for a function can significantly improve the performance of code generation tools in generating accurate code. It is important for practitioners to assign clear and descriptive names to functions. However, our findings suggest that providing thorough explanations for functions is even more critical than giving them meaningful names. Ideally, both practices should be employed to produce the most accurate and valid code possible.
        \end{minipage}}
        \medskip

        \begin{figure}[h]
\begin{lstlisting}[caption={Generated Code for the Example Problem (ID: 161)}, label={prompt-161-copilot},  ]
def solve(s):
    """You are given a string s. if s[i] is a letter, reverse its case from lower to upper or vise versa, otherwise keep it as it is. If the string contains no letters, reverse the string. The function should return the resulted string.
    Examples
    solve("1234") = "4321"
    solve("ab") = "AB"
    solve("#a@C") = "#A@c"
    """


#Result of the old version of Github Copilot
    return ''.join(c.lower() if c.isupper() else c.upper() for c in s) or s[::-1]

#Result of the new version of Github Copilot
    if not any(c.isalpha() for c in s):
        return s[::-1]
    return ''.join(c.swapcase() if c.isalpha() else c for c in s)
\end{lstlisting}
        \end{figure}

        \subsection{Evaluation of Code Generation Tools Over Time (RQ4)}\label{Discussion-Evaluation-of-Tools-Over-Time}

 \begin{figure*}
           \includegraphics[scale=0.70]{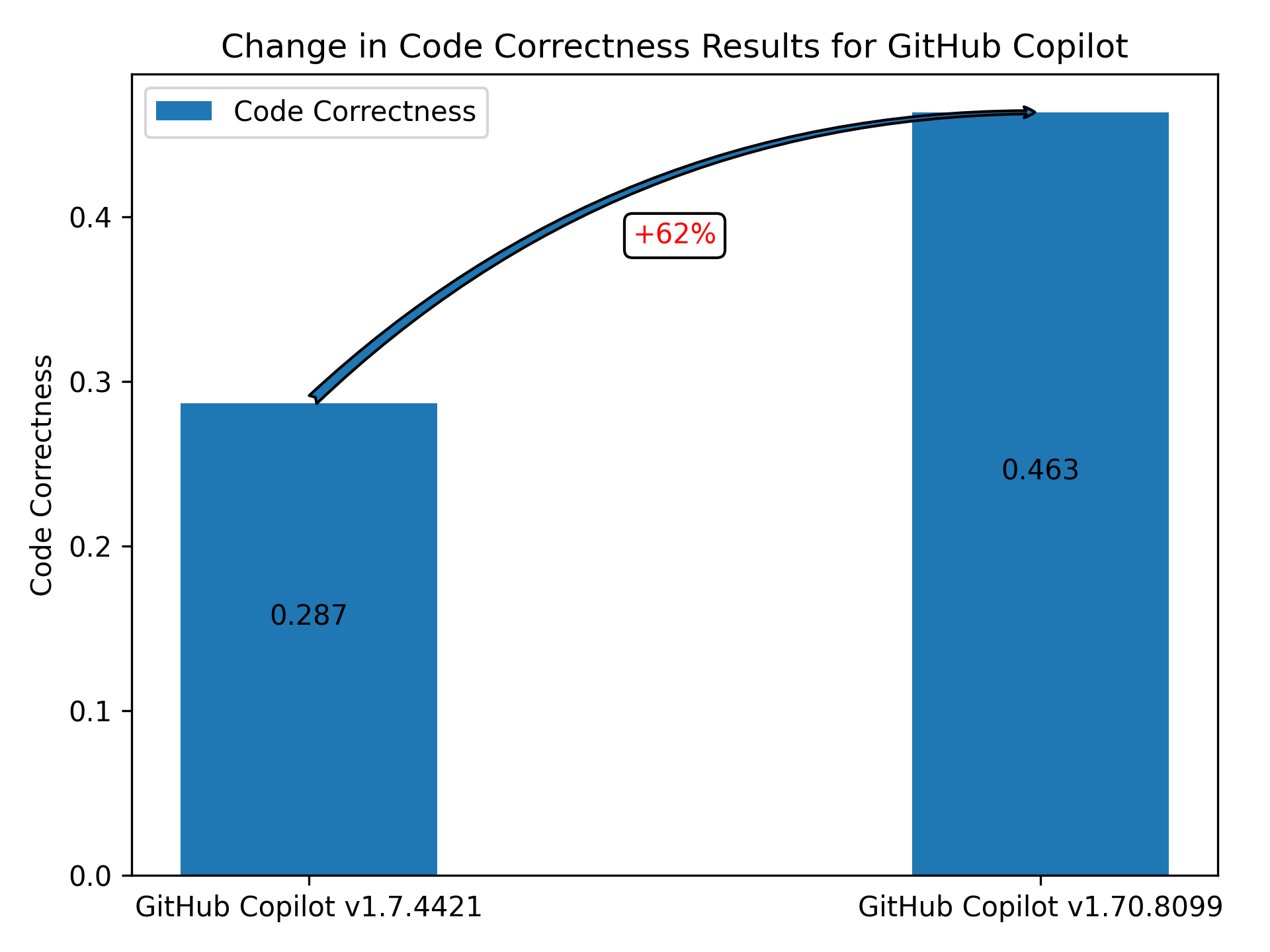}
            \caption{GitHub Copilot Improvement in Code Correctness}
            \label{experimental_workflow_correctness_copilot}
        \end{figure*}

        \begin{figure*}
            \includegraphics[scale=0.70]{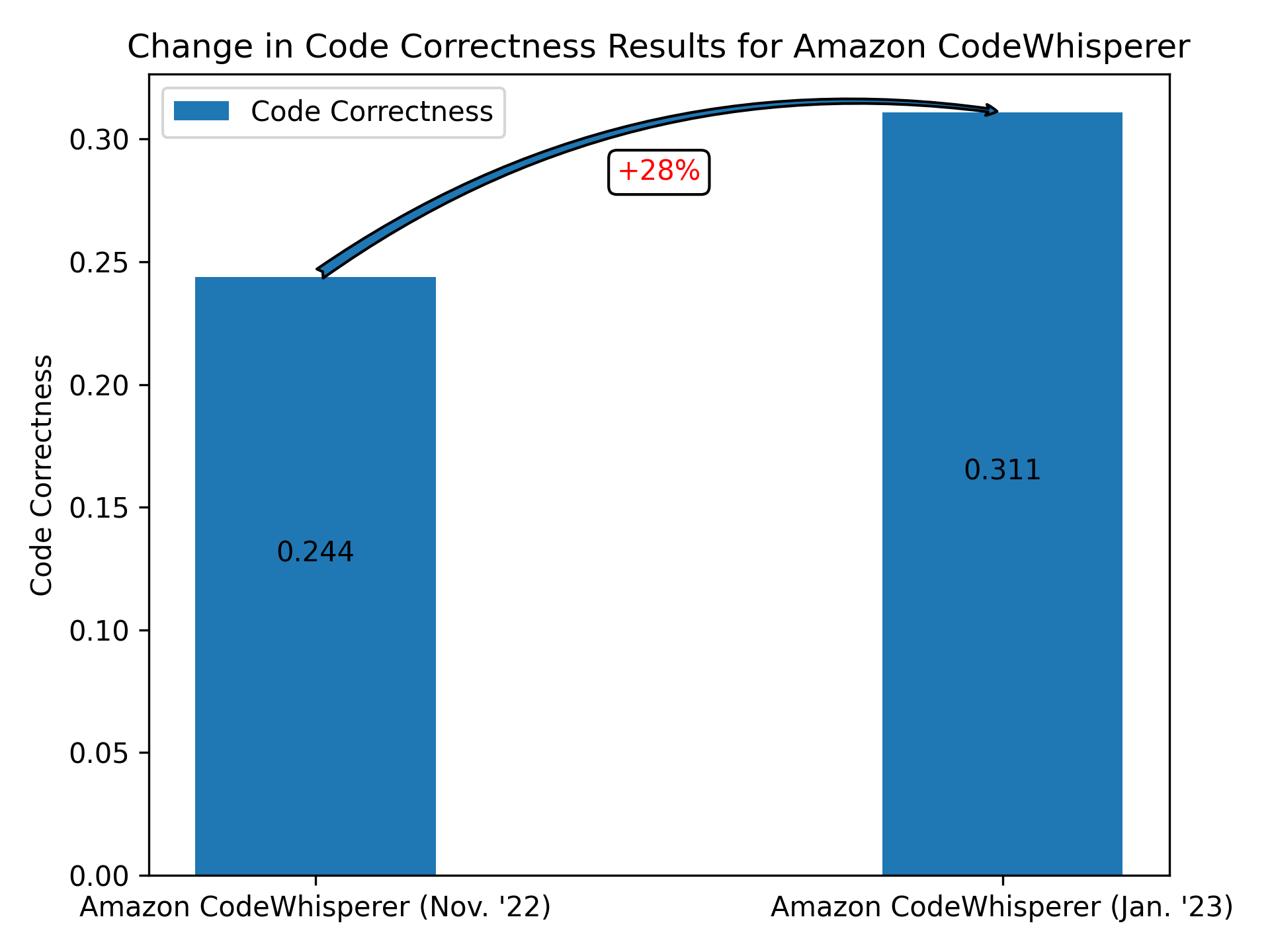}
            \caption{Amazon CodeWhisperer Improvement in Code Correctness}
            \label{experimental_workflow_correctness_whisperer}
        \end{figure*}

                \begin{figure*}
            \includegraphics[scale=0.70]{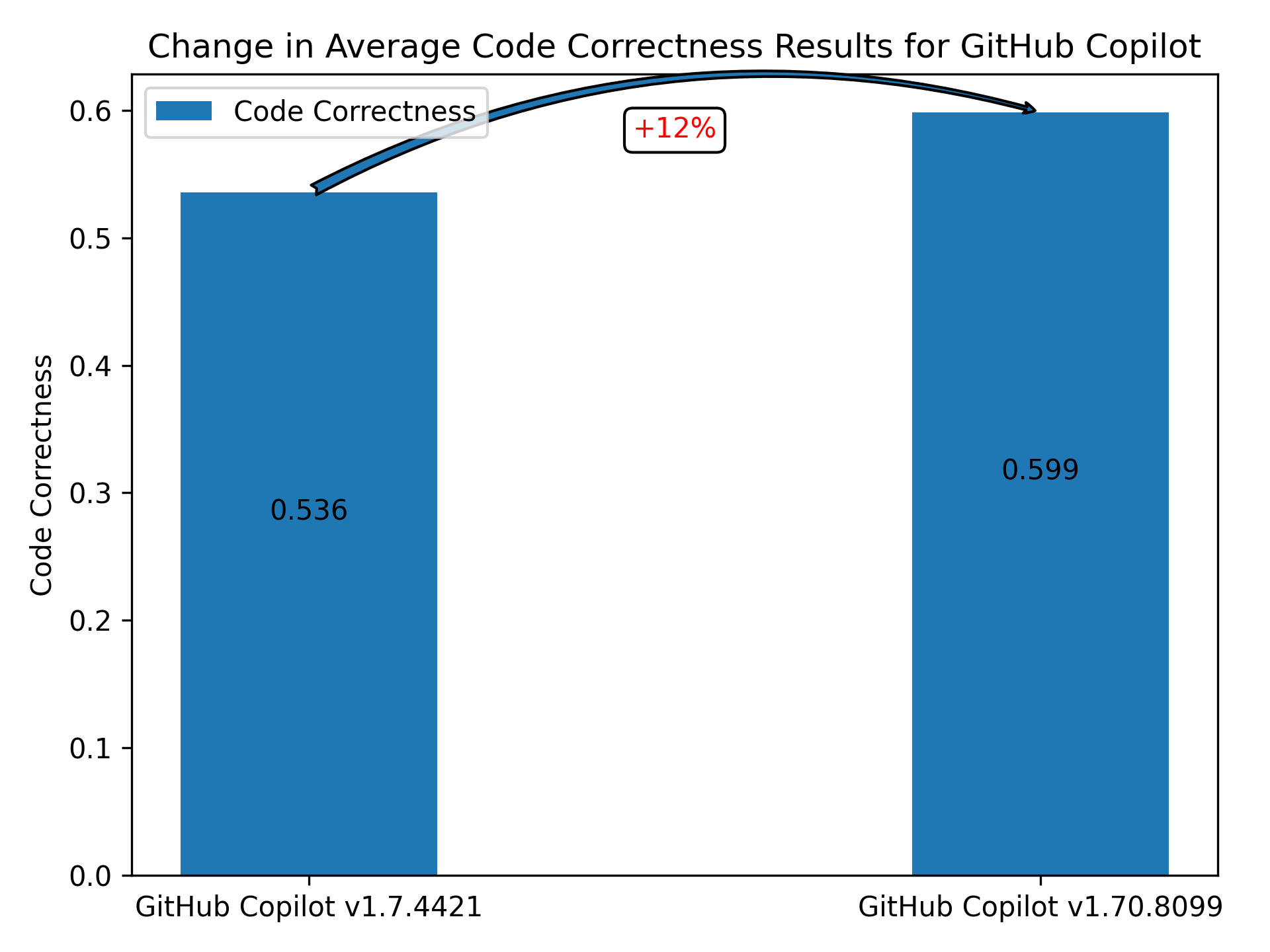}
            \caption{GitHub Copilot Improvement in Average Code Correctness}
            \label{experimental_workflow_avrg_correctness_copilot}
        \end{figure*}

        \begin{figure*}
            \includegraphics[scale=0.70]{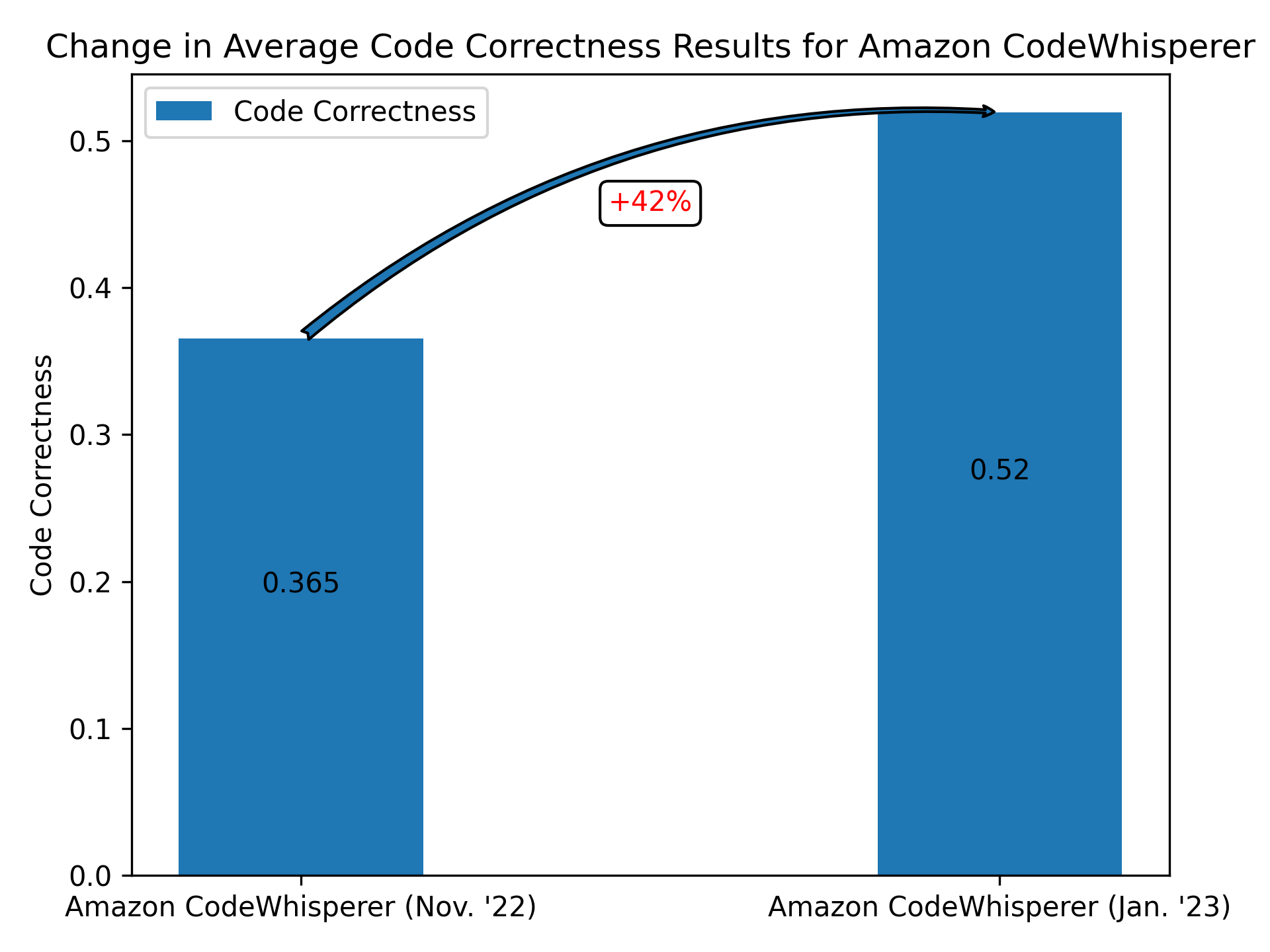}
            \caption{Amazon CodeWhisperer Improvement in Average Code Correctness}
            \label{experimental_workflow_avrg_correctness_whisperer}
        \end{figure*}
      
        The results mentioned in Section \ref{Results-Evaluation-of-Tools-Over-Time} demonstrate that both GitHub Copilot and Amazon CodeWhisperer increased the number of correct code suggestions on their new versions. As it can be observed from Figure \ref{experimental_workflow_correctness_copilot} and \ref{experimental_workflow_avrg_correctness_copilot}, GitHub Copilot v1.70.8099 had passed suggestions for 46.3\% unit tests and average correctness for 59.9\% of all problems, while GitHub Copilot v1.7.4421 had passed suggestions for 28.7\%  the unit tests and average correctness for 53.6\% of all problems. This data indicates that there is a notable improvement of 62\% in GitHub Copilot's performance in terms of passed suggestions for unit tests and there is an improvement of 12\% in the average correctness results. Figure  \ref{experimental_workflow_correctness_whisperer} - \ref{experimental_workflow_avrg_correctness_whisperer} provides that Amazon CodeWhisperer had passed solutions for 24.4\% of all unit tests  and average correctness for 36.5\% of all problems in November 2022, while it provided passed solutions for 31.1\% of all problems and average correctness for 52\% of all problems in January 2023. These results demonstrate an improvement of 28\% in terms of passed suggestions of unit tests and there is an improvement of 42\% in the average correctness results of Amazon CodeWhisperer.

  The new version of GitHub Copilot enhanced the correctness of 65\% of the partially correct recommendations from the previous version, according to our analysis of the findings. We also noticed that 9 out of the 33 incorrect answers from the previous version of GitHub Copilot were better in the new version. GitHub Copilot generally suggested code that was close to the prior version. Minor but substantial changes were made, though, to enhance the recommendation's accuracy.  To illustrate, The Listing \ref{prompt-161-copilot} shows the code recommendations of two versions of GitHub Copilot for the same problem. The problem \#161 in the experiment where the input for the function is a string, expected to swap the case of the letters in the string if there are any. If there are no letters in the string, the function should only return the reversed string. The previous version of GitHub Copilot offered an incorrect recommendation for this problem since the expression related to reversing the string was not located appropriately within the function. In the new version of GitHub Copilot, we observed that although its suggestion was generally very similar to the previous one, it placed the swapping expression in the correct location in the function and this modification caused an increase in correctness from 75\% to 100\%.

Additionally, we performed a comparison of the outcomes between the two different Amazon CodeWhisperer versions. According to our research, 50\% of the partly correct suggestions made by the prior version of Amazon CodeWhisperer were made more correctly by the new version. In addition, we discovered that 46 out of the 74 incorrect suggestions made by the prior version of Amazon CodeWhisperer were made correctly by the new version. In most cases where the new version of Amazon CodeWhisperer improved upon the previous version, we observed that the previous version frequently returned default values such as empty list, zero, false, or empty string. Although these recommendations sometimes produced correct results with some unit tests, they resulted in a low rate of partial correctness. With the updated Amazon CodeWhisperer, this problem was largely resolved, and the updated version offered more logical and precise answers. 
        
Additionally, we noticed that 10 out of 14 incorrect answers from the previous version of GitHub Copilot were removed in the new version. These incorrect answers were typically the result of syntactical errors, excess of maximum recursion depth, or the usage of modules that are not imported. For example, the Listing \ref{prompt-123-copilot} shows that the old version of GitHub Copilot recommended an unsuccessful recursion function without a base case for the problem \#123, which leads to a compilation error. As a result of introducing a base case to the code, the new version of GitHub Copilot was able to fix this error completely. As another example, while the old version of GitHub Copilot did not add import the `reduce' module, the new version of GitHub Copilot recommended a different solution without using the `reduce' module, which resulted in 100\% correctness. However, we found that almost none of the invalid suggestions made by the earlier version of Amazon CodeWhisperer changed their status in the new version when we analyzed the validity rate for both versions.

\begin{figure}[h]
\begin{lstlisting}[caption={Generated Code for the Example Problem (ID: 123)}, label={prompt-123-copilot},  ]
def get_odd_collatz(n):
    """
    Given a positive integer n, return a sorted list that has the odd numbers in collatz sequence. The Collatz conjecture is a conjecture in mathematics that concerns a sequence defined as follows: start with any positive integer n. Then each term is obtained from the previous term as follows: if the previous term is even, the next term is one half of the previous term. If the previous term is odd, the next term is 3 times the previous term plus 1. The conjecture is that no matter what value of n, the sequence will always reach 1.
    Note: 
        1. Collatz(1) is [1]. 2. returned list sorted in increasing order.
    For example:
    get_odd_collatz(5) returns [1, 5] # The collatz sequence for 5 is [5, 16, 8, 4, 2, 1], so the odd numbers are only 1, and 5.
    """


    #Result of old version of GitHub Copilot in the original experiment
    return [n] + get_odd_collatz(3 * n + 1) if n % 2 == 1 else get_odd_collatz(n // 2)

    #Result of new version of GitHub Copilot in the original experiment
    return sorted([i for i in collatz(n) if i%2])
def collatz(n):
    if n == 1:
        return [1]
    if n%2:
        return [n] + collatz(3*n+1)
    return [n] + collatz(n//2)
\end{lstlisting}
\end{figure}

On the other hand, our research showed that 17 out of the 47 correct answers provided by the previous version of GitHub Copilot had incorrect or only partly correct answers in the new version. The outcomes also showed that of the 84 partial correct answers in the old GitHub Copilot version, 19 of them had reduced percentages of correctness in the new GitHub Copilot version. When we examined the answers from the two versions of GitHub Copilot, we found that the new version of GitHub Copilot frequently recommended ´pass', which reduced the correctness of the answer. We observed that GitHub Copilot sometimes answered the problem nearly correctly, and sometimes it only answered as `pass' to the same problem. These issues may be related to the undetermined nature of GitHub Copilot. We also noticed that some problems were resolved in the previous version of GitHub Copilot with the correct solutions. Still, in the new version of GitHub Copilot, these problems resulted in compilation errors, which were generally because of forgetting to add an import to the modules.

Our research also showed that while the earlier version of Amazon CodeWhisperer offered entirely correct suggestions, the new version provided partial or incorrect answers for 16 of the 40 problems. Additionally, our research revealed that the new version of Amazon CodeWhisperer had reduced correctness rates for 20 of the 50 partly correct solutions offered by the prior version. We noticed that the following were the main reasons why code recommendations in the new edition were unsuccessful: Firstly, the tool occasionally suggested using a pass statement only, which is an incorrect response to the problem at hand. Second, as was already stated, it occasionally returned default numbers that did not fulfill the required solution. Thirdly, it proposed a single-line return expression that was complicated and often failed to produce correct results. For instance, Listing \ref{prompt-155-codeWhisperer} presents that problem \#155 wanted the function should take a number as an input parameter, and the function should return a tuple that has the number of even and odd digits, respectively. The new Amazon CodeWhisperer version for this problem produced a single-line but complicated return statement which gave a syntactic error for negative numbers. Despite having a long solution, the previous version of Amazon CodeWhisperer had an 85\% correctness rate.

\begin{figure}[h]
\begin{lstlisting}[caption={Generated Code for the Example Problem (ID: 155)}, label={prompt-155-codeWhisperer},  ]
def even_odd_count(num):
    """Given an integer. return a tuple that has the number of even and odd digits respectively.
     Example:
        even_odd_count(-12) ==> (1, 1)
        even_odd_count(123) ==> (1, 2)
    """

 
    #Result of old version of Amazon CodeWhisperer in the original experiment
    even = 0
    odd = 0
    if num < 0:
        num = -num
    while num > 0:
        if num % 2 == 0:
            even += 1
        else:
            odd += 1
        num = num // 10
    return even, odd

    #Result of new version of Amazon CodeWhisperer in the original experiment
    return (len([i for i in str(num) if int(i)%2==0]), len([i for i in str(num) if int(i)%2]))
\end{lstlisting}
\end{figure}
       
         \medskip
        \fbox{\begin{minipage}{33em}
  GitHub Copilot's new version had 62\% more passed-unit tests than its older version. Similarly, Amazon CodeWhisperer's updated version resulted in 28\% more passed-unit tests than its previous version, suggesting that both tools have notable improvements.
        \end{minipage}}
        \medskip

    \section{Threats to Validity}\label{Threats-to-Validity}

        \subsection{Conclusion Validity}
            \begin{itemize}[leftmargin=0cm]              
                \item[] \textbf{Trivial Solutions:} In some problems, code generators generated solutions to return simple statements like empty arrays or Boolean values. In this case, if there are test cases related to the problem, where such expressions are the desired output, those test cases pass by chance without any algorithm generated for the problem.\\

                \item[] \textbf{Number of test cases:} The varying amount of test cases for the dataset may introduce a threat to our experiment. On average there are 7.7 test cases for each problem in the HumanEval dataset \cite{copilot2021}. Having broader test cases, both for the amount and the scope can be important. By extending the test cases, any potential corner case that could be missed may be covered. This can be critical especially when some corner cases for a given problem are not involved. We plan to improve the test cases both in quantity and quality in our future work.\\

                \item[] \textbf{SonarQube:} We used the SonarQube code inspector to obtain results for our code security, code maintainability, and code reliability metrics. But in our results for them, there was scarce information about the possible vulnerabilities for the generated code in each sample, which would be discovered by SonarQube. We believe that due to the extent of the problems that are contained in the HumanEval Dataset, the solutions to those problems consist of a small number of lines. The case could also be observed in the canonical solutions provided with each question. 
            \end{itemize}
        
        \subsection{Internal Validity}
            \begin{itemize}[leftmargin=0cm]
                \item[] \textbf{One-shot code generation:} While generating code with the code generators, we used the function names, parameters, the corresponding docstring containing an explanation of the function, and a few instances of tests for that function. Furthermore, we did not write any code to provide additional information to the code generators which would clarify more what our intent for that particular problem is. Therefore, in most cases, the success rate could be increased if we have given hints as code snippets to the code generators.\\
          
                \item[] \textbf{Reproduction of the Generations:} While conducting our experiments, we observed that the code generators had a nondeterministic characteristic, hence they were generating different outputs for the same input in different trials. We paid great attention not to include different outputs for the same input by generating code for our problems in one iteration, and saving the generated code, then conducting our evaluation on the saved code. Given that the code generators have a dynamic characteristic that the underlying LLM of the tools is being retrained, our results might not be fully replicated given our experimental setup and input.\\
        
                \item[] \textbf{Code Generation Methods:} With GitHub Copilot and Amazon CodeWhisperer, one can use two different approaches to generate code. For GitHub Copilot, the first one happens automatically as a programmer proceeds to write code, GitHub Copilot suggests code snippets that might fit into that context. In the other approach, whenever the programmer wants to generate code, they press the `ctrl + enter' key combination to see up to 10 code generations GitHub Copilot produces. Similar to GitHub Copilot, the default approach for Amazon CodeWhisperer is also automatically generating code when prompted. The other approach in Amazon CodeWhisperer is to use the `option + c' (Mac) / `alt + c' (PC) key combination. In our experiment, we chose the first approach whenever possible, otherwise, we implemented the second approach and selected the suggestion at the top of the list. There were some problems where the generators failed to generate any code after we entered the next line (after the user presses the `enter' key and continues from the next line). Therefore, we had to apply the key combinations to see the solutions, and we were able to obtain code generations for all problems. As we had to use two different methods for code generation, we stated our practice as a possible factor to reduce the validity of our study. \\
        
                On a further note, we also want to state the difference between the solutions that are automatically generated, and the ones shown when the key combinations are applied. We observed that for the same context, two methods yield different results. Therefore, if in both methods, the code is generated, choosing different methods for a set of problems may introduce possible invalidity to a study. Hence, we tried to be as consistent as possible in our experiment by avoiding the latter method whenever possible.\\
            
                \item[] \textbf{Block and Line-by-Line Generation:} For GitHub Copilot and Amazon CodeWhisperer, for most of the cases, they managed to generate the solution of a given function as bulk, but there were cases, where we had to generate the solution line-by-line. As we had no control over how these tools would generate the code, we had to accept the method they would choose for a particular problem. We state these cases, as in line-by-line suggestion, the previously generated lines might have an effect on the next line to be generated, whereas in the first case, code is generated at once as a bulk. We have not experienced this problem with ChatGPT, since it always generated the whole function in a single interaction.\\
        
                \item[] \textbf{Versions of the Generators:} While conducting our experiment, the latest version of GitHub Copilot was \textit{v1.70.8099} and of ChatGPT was \textit{9 Jan '23}. As explained earlier, Amazon does not keep the version of their CodeWhisperer, therefore we can only provide the time when we conducted our experiment, which was January 2023. For any possible later evaluations with the same experimental setup, the results might be different, which we have proved with RQ4.\\

                \item[] \textbf{Interval for Improvement of the Code Generators:} As mentioned in Section \ref{Methodology-Evaluation-of-Code-Generation-Tools-Over-Time}, we have made duplicate experiments on GitHub Copilot and Amazon CodeWhisperer to evaluate how these tools have improved over time, in terms of code correctness. However, the intervals between the two experiments for each generator are not the same. In other words, the time difference between the experiments we conducted for GitHub Copilot was 13 months; however, this difference was two months for Amazon CodeWhisperer. Hence, the room for improvement for these tools is not the same.\\

                \item[] \textbf{Prompting ChatGPT:} GitHub Copilot and Amazon CodeWhisperer are tools that are integrated into an IDE. Therefore when we provided input to these tools, the tools generated code directly. However, we had to give an explanation to ChatGPT that our inputs should be used to generate code. We explained in a the following sentence ``Generate code using the prompts I will provide" to ChatGPT for code generation. Then we inputted the prompts one-by-one in the same chat window.
            \end{itemize}
          
        \subsection{Construct Validity}
            \begin{itemize}[leftmargin=0cm]          
                \item[] \textbf{Metrics:} As explained in Section \ref{Methodology-Our-Metrics}, we have evaluated the generated code using the Code Validity, Correctness, Security, Maintainability, and Reliability metrics. However, we are aware that we could have used additional metrics, such as Readability, Cyclomatic Complexity, and Reusability to analyze the generated code.
            \end{itemize}
            
        \subsection{External Validity}
            \begin{itemize}[leftmargin=0cm] 
                \item[]\textbf{Problem Coverage:} For our experiment, we evaluated the generated solutions for 164 different problems, contained in the HumanEval dataset. In the HumanEval dataset, the subjects of the problems include algorithms, simple mathematics, reasoning, and language comprehension \cite{copilot2021}. For better and more insightful results, the number of problems can be increased, and the comprehension of the problems could be broader. For instance, in the experimental setup proposed by \cite{inIDE_code} for their code generation and retrieval tool, the scope of the problems consists of basic Python, file, OS, web scraping, web server \& client, data analysis \& ML, and data visualization. Such topics could be included in our dataset to both broaden the comprehension and increase the number of our problems. We consider this task as future work for our study. \\

                \item[] \textbf{Dependency on the HumanEval Dataset:} As we explained in Section \ref{Methodology-HumanEval-Dataset}, we have used the HumanEval Benchmark Dataset for our experiment. Since we have only used this dataset, it followed that we were limited to the Python programming language. Hence, our experiment can reflect the code generation performance of the generators in regard to this language. \\

                \item[] \textbf{IDE Dependency:} For code generation using GitHub Copilot and Amazon CodeWhisperer, we used the Visual Studio Code IDE. In Table \ref{tab:comparison-table}, we show the available IDEs that these tools are available on. Since we tried the code generation only on a single IDE, there might be IDE-dependent results; using the other mentioned IDEs might have yielded different results.
            \end{itemize}

    \section{Related Work}\label{Related-Work}
    
    In the last few years, code generation has attracted attention from researchers such as \cite{codegen, recode, treegen, Lyu2021}. In this study, we compare some of the prevalent code generators: GitHub Copilot, Amazon CodeWhisperer, and ChatGPT,  concerning their code generation capabilities, and find the strengths and shortcomings of the tools by looking at the code quality metrics: Code Validity, Code Correctness, Code Reliability, Code Security, and Code Maintainability. When we investigated the studies similar to ours, we found more related studies about GitHub Copilot than other AI-based code generation tools. The reason is that GitHub Copilot was the first created tool among the code generators we evaluated and there was more time for it to be evaluated by researchers.
    
    The underlying model of GitHub Copilot, Codex, is externally developed by OpenAI and employed by GitHub. Some earlier versions of the current Codex model used by GitHub Copilot were evaluated by \cite{copilot2021}. The Codex model relies on GPT models that OpenAI previously developed for natural language generation. The public code available on GitHub was used here while fine-tuning the model to implement the code recognition and generation capabilities. This model can recognize other elements such as function signatures, code comments, etc. and it can use such elements as inputs and generate related outputs. They found that a success rate of 70.2\% could be reached in terms of code correctness, by generating 100 solutions for each problem and choosing the most successful one among them. The success rate was only 28.8\% for the case with one solution per problem, which is consistent with our initial results in \cite{yetistiren2022assessing}. In this research, we also examined code reliability, maintainability, and security to provide a detailed form of this evaluation. We evaluated the three main code-generation tools as well. Additionally, in order to broaden the scope of our investigation, we modified the HumanEval dataset by replacing real function names with the dummy name `foo' and then produced new sets of results.
    
    
    There are also empirical studies similar to ours, conducted to evaluate GitHub Copilot. We list the available studies in the following. 
    
    One such study is conducted by  \cite{programming_copilot} in which the code correctness of GitHub Copilot is evaluated, and the tool is contrasted to the automatic program generators having the Genetic Programming (GP) architecture. They found that there is not a significant difference between the two approaches on the benchmark problems; however, the program synthesis approaches are not sufficient in supporting programmers compared to GitHub Copilot. 
    
    An evaluation of GitHub Copilot in terms of the security of the generated programs was implemented by \cite{asleep}. They evaluated the vulnerabilities in the code generated by Copilot. It was determined that 40\% of generated programs were vulnerable. These results differed from our Code Security results; we believe that the characteristics of our dataset caused the difference between the two studies.
    
    Another study discusses the effects of GitHub Copilot by conducting a within-subjects user study \cite{copilot-user-exp}. It was found that GitHub Copilot did not cause a significant improvement in terms of speed and success rate. However, it was stated that most participants preferred to use Copilot in daily programming tasks since it saved the effort for the basic tasks.  
    
     \cite{nguyen} evaluated GitHub Copilot using 33 different LeetCode questions and four different programming languages (Python, Java, JavaScript, and C). Their evaluation includes code correctness and code understandability for the generated code. They evaluated code correctness by measuring the ratio of passed tests for each question, which is a similar approach to our study. Code understandability was measured by two different metrics, which are cognitive and cyclomatic complexity. In terms of code correctness, Java had the highest (57\%) and JavaScript had the lowest (27\%) score. For code understandability, they determined that there was no statistical significance between the programming languages.

     \cite{mastropaolo2023robustness} presented an empirical study that focuses on the effect of semantic-preserving changes in the natural language on the generated code function of GitHub Copilot. For this purpose, \cite{mastropaolo2023robustness} provided 892 non-trivial Java method descriptions to GitHub Copilot. Firstly, they used the original descriptions of methods and asked GitHub Copilot to generate them. Secondly, they paraphrased descriptions manually. Thirdly, they paraphrased descriptions using automated paraphrasing tools. After GitHub Copilot generated all of the methods according to their descriptions, they found that in ~46\% of cases, semantically equivalent but different method descriptions resulted in different code recommendations. Moreover, they observed that some code recommendations were correct with only one of the semantically equivalent descriptions as input.

     ChatGPT is the other code generator that we have chosen for our study.
     Since ChatGPT is released recently, there are only a few studies similar to our work. These studies are in the following.

     In order to analyze the bug fixing performance of ChatGPT, \cite{sobania2023analysis} evaluated ChatGPT on the standard bug fixing benchmark set, QuixBugs, and compared these results with CoCoNut, Codex, and standard APR approaches. They found that ChatGPT had a similar performance to Codex and its performance was much better than standard APR approaches. When \cite{sobania2023analysis} used the dialogue option of ChatGPT and gave ChatGPT more information about the bug, they found that ChatGPT gave an overall success rate of 77.5\%. Then, they concluded that although ChatGPT had an outstanding performance, it required mental cost to verify ChatGPT answers.

The possible integration of ChatGPT into a well-known software testing curriculum is covered in another research. \cite{jalil2023chatgpt} requested that ChatGPT respond to typical software testing questions. They discovered that ChatGPT could offer correct or partly correct responses in ~44\% of the cases and correct or partially correct explanations of answers in ~57\% of the cases. As a result, they noticed that ChatGPT failed a course on software testing. Furthermore, ChatGPT was a poor judge of its correctness. It is therefore uncertain how much benefit ChatGPT might offer to students.
    
Additionally, we have selected Amazon CodeWhisperer as the code generation tool for our study's evaluation. There are no studies about Amazon CodeWhisperer that are comparable to our research because it is a relatively new tool,  like the case with ChatGPT.
    
In general, most of the associated works assessed the code quality produced by GitHub Copilot or OpenAI's Codex model. Except for the studies of \cite{asleep}, where the emphasis is code security, and \cite{copilot-user-exp}, where the work primarily focuses on the practical usage performance of GitHub Copilot, the majority of the studies concentrated on the assessment of code correctness. On the other hand, there aren't nearly enough studies about ChatGPT and Amazon CodeWhisperer that are comparable to ours. The bug-fixing performance of ChatGPT \citep{sobania2023analysis} and the possible applicability of ChatGPT to a well-known software testing program \citep{jalil2023chatgpt} were the main topics of related research about ChatGPT that we discovered.
    
To the best of our knowledge, this is the first study comparing code correctness, code validity, security, maintainability, and dependability between GitHub Copilot, Amazon CodeWhisperer, and ChatGPT. In this regard, we think that our methods and findings will contribute to current research on the capabilities of these common tools and other code-generation tools.

    \section{Conclusion}
    \label{Conclusion}
    In our study, we compared three code generation tools: GitHub Copilot, Amazon CodeWhisperer, and ChatGPT. We evaluated the quality of the generated code in terms of correctness, validity, reliability, security, and maintainability. Our results show that ChatGPT, in its original setup, had the highest success rate among the evaluated code generation tools. Specifically, it was able to generate correct code solutions for 65.2\% of the problems in the HumanEval problem dataset. It also produced partially correct solutions for 22.6\% of the problems and incorrect solutions for 12.2\% of the problems.

    In terms of code maintainability, we found repeated types of code smells among the code generation tools, and we compiled a list of the code smells that we encountered. If the solution to the problem contained any smells, the average time to eliminate them was 9.1 minutes for GitHub Copilot, 5.6 minutes for Amazon CodeWhisperer, and 8.9 minutes for ChatGPT.

    To evaluate the impact of input parameters' quality on the three major code generation tools, we conducted an assessment of providing only function names and parameters. Our findings revealed that compared to their initial setup, all three code-generation tools had lower percentages of correct answers. ChatGPT and GitHub Copilot achieved the best and most comparable outcomes, with correct answers for 20\%-22\% of the problems, partially correct answers for 26\%-27\% of the problems, and incorrect answers for 50\%-53\% of the problems.

    We also investigated the effect of dummy function names on the success of code generation tools. Our results showed that ChatGPT had the highest percentage of correct solutions among the three tools, with 61.6\% of the problems examined generating correct solutions. ChatGPT also generated partially correct solutions for 25.6\% of the problems and incorrect solutions for 12.8\% of the problems.

    Based on our findings, ChatGPT was the most successful tool, whereas Amazon CodeWhisperer was the least successful. We also found that providing an accurate and clear problem description was essential for the success of code generation tools, as we observed that all tools performed worse when we eliminated docstrings from the input. Moreover, we observed that both Amazon CodeWhisperer and GitHub Copilot are improving rapidly, suggesting their potential for various coding tasks in the future.

    In summary, our study has made several contributions to the understanding of code generation tools:
    
    \begin{itemize}
        \item We conducted a comparative analysis of GitHub Copilot, Amazon CodeWhisperer, and ChatGPT and provided a comprehensive comparison table (Table \ref{tab:comparison-table}) of their features.
        \item We evaluated the code generation capabilities of these tools using the HumanEval dataset and proposed a pipeline to assess the quality of the generated code. 
        \item We analyzed the performance improvements of the new version of GitHub Copilot compared to the previous version and observed the improvement of Amazon CodeWhisperer between November 2022 and January 2023.
        \item We highlighted the importance of providing accurate and clear problem descriptions for code generation tools to improve their performance.
    \end{itemize}

    Overall, our study contributes to the development of code generation tools and provides insights into their potential for various coding tasks in the future.
    
    \begin{itemize}[leftmargin=0cm]
        \item[]\textbf{Data Availability:} Our data yielded from this study and the code that was utilized can be found in our repository at \href{https://github.com/mirayayerdem/Github-Copilot-Amazon-Whisperer-ChatGPT}{https://github.com/mirayayerdem/Github-Copilot-Amazon-Whisperer-ChatGPT}.\\
        \item[]\textbf{Conflict of Interests:} The authors declare no conflicts of interest in relation to this article.
    \end{itemize}

    \bibliographystyle{spbasic}
    \bibliography{references}
    
\end{document}